\documentclass[journal, compsoc]{IEEEtran} 

\pdfoutput=1

\usepackage{blindtext}
\usepackage{graphicx}
\usepackage[utf8]{inputenc}
\usepackage{hyperref} 
\usepackage{amssymb} 
\usepackage{amsmath} 
\usepackage{subfig} 
\usepackage{natbib}
\setcitestyle{authoryear}

\makeatletter
\def\@seccntformatinl#1{\csname the#1dis\endcsname\hskip 1em\relax}
\makeatother

\begin{document}
%
\title{Light interception modelling using unstructured LiDAR data in avocado orchards}

\author{Fred~Westling*,
        James~Underwood,
        Samuel~Örn
\thanks{S. Örn is with Linköping University}
\thanks{F. Westling and J. Underwood are with the University of Sydney}
\thanks{*Correspondence: f.westling@acfr.usyd.edu.au}
}

\maketitle

\begin{abstract}
In commercial fruit farming, managing the light distribution through canopies is important because the amount and distribution of solar energy that is harvested by each tree impacts the production of fruit quantity and quality. It is therefore an important characteristic to measure and ultimately to control with pruning. We present a solar-geometric model to estimate light interception in individual avocado (Persea americana) trees, that is designed to scale to whole-orchard scanning, ultimately to inform pruning decisions.

The geometry of individual trees was measured using LiDAR and represented by point clouds. A discrete energy distribution model of the hemispherical sky was synthesised using public weather records. The light from each sky node was then ray traced, applying a radiation absorption model where rays pass the point cloud representation of the tree.

The model was validated using ceptometer energy measurements at the canopy floor, and model parameters were optimised by analysing the error between modelled and measured energies. The model was shown to perform well qualitatively well through visual comparison with tree shadows in photographs, and quantitatively well with $R^2 = 0.854$, suggesting it is suitable to use in the context of agricultural decision support systems, in future work.

\end{abstract}


\IEEEpeerreviewmaketitle

\section{Introduction}

For commercial fruit trees, the total light available to the tree and its distribution throughout the canopy is a primary factor leading to the production of quality fruit as explained by \cite{McFadyen2004}, as well as fruit characteristics including dry weight and oil concentration (\cite{connor2016relationships}).  Certain crops require a relatively large amount of energy to meet minimal acceptable quality standards, like those described for avocado by \cite{Lee1983}, yet particularly dense trees can have a high total light interception but inadequate light distribution, preventing significant parts of the tree from contributing to yield.  Consequently, methods for estimating the magnitude and distribution of light throughout canopies are of interest, both for yield estimation and to support crop management decisions.

The motivation of this paper is to create a model that associates the distribution of light to the geometrical structure of a tree, so that future work can use this to recommend pruning actions that result in optimal light distribution.

As the total light used by a tree for photosynthesis during a growing period is difficult to observe directly, it is commonly inferred from the amount of light intercepted by the tree at instantaneous measurement times as discussed by~\cite{Cherbiy-Hoffmann2013,Cherbiy-Hoffmann2012}.

Methods of modelling light interception have long been used to inform orchard decision processes with regards to optimal spacing and tree shape.  ~\cite{charles1982physiological} describes how to analytically estimate light interception using a simplified model which incorporates consideration of canopy geometry, varying foliage density and leaf orientation, but uses minimal complexity due to the lack of computing power available at the time.  
More recent models promote a greater understanding of crop growth.  Functional-Structural Plant Modelling (FSPM) like that described by \cite{White2012} can be used to generate tree models with sufficient resolution to apply high-fidelity light environment modelling like QuasiMC presented by \cite{cieslak2007quasi}, which can also be extended to provide high-quality information regarding the plant's response to changing management practices.  \cite{Massonnet2008} demonstrated virtual systems for simulating respiration, transpiration and photosynthesis in apple trees, allowing growers to see what effect a particular pruning strategy might have on a typical tree. 
These methods require high resolution digitisations of the trees, which are difficult to obtain for physical trees using sensors but can be achieved by simulating the growth of virtual trees.

Lower resolution methods exist to measure the energy absorption characteristics of real-world trees.  Ceptometers, which measure the amount of Photosynthetically Active Radiation (PAR) at the sensor, can be used to estimate light interception by subtracting simultaneous above and below canopy measurements.  In order to achieve an accurate estimation, measurements for any given tree must be taken multiple times in different light conditions and at many locations under the tree to reduce the effect of spatial noise due to patchy light, as described by \cite{ibell2015preliminary}.  This method is sufficiently useful that it is often employed despite the intensive labour and time required, so there is interest in faster and easier methods.

Light interception can alternatively be estimated using plant characteristics, including Leaf Area Index (LAI) which measures the leaf area per unit ground area.  An early method used mechanical pin frames (\cite{wilson1963estimation}) but more recently sensing devices such as Pocket LAI may be used (\cite{confalonieri2013development,francone2014comparison}).  Here, a camera on a mobile phone is used to compute the gap fraction of a plant, which means how much of the sky is visible through the foliage when the plant is viewed from the ground at a certain angle.

Alternative methods to analyse individual trees can be used if accurate geometric model of the trees are available, and recent advances in Light Detection and Ranging (LiDAR) enable quick and accurate capture of such models at a low cost (\cite{rosell2012review}).  LiDAR systems mounted on ground vehicles can be used to measure tree parameters like height, volume and leaf area (\cite{Nielsen2009,Sanz-Cortiella2011,underwood2016mapping}), while statically mounted terrestrial LiDAR like that used by \cite{Kato2009} can generate higher quality models at the cost of time and scalability.

 \cite{hagstrom2011line} describe a LiDAR-based approach for estimating tree characteristics where the opacity of voxelised data is computed by calculating how LiDAR beams pass through each voxel, compared to how much is reflected, to estimate gap fraction similarly to the Pocket LAI method.  While LiDAR scanning is limited in resolution, certain information beyond raw geometry can be deduced.  \cite{Ma2016,Ma2016a} present a method for segmenting terrestrial LiDAR point clouds into photosynthetic and non-photosynthetic components with 91\% accuracy using purely geometric data, and further demonstrate that they can calculate the woody-to-total-area ratio of individual trees, which can be used to estimate the LAI and improve radiosity simulations.

Further analysis can be applied by employing a combination of modelling and measurement.  An early attempt using Silhouette to Total Area Ratio (STAR) to measure light interception is presented by \cite{sinoquet2005foliage}.  In this method, the ratio of silhouette size to total leaf area was calculated for digitised and simulated trees integrated over a discretised sky model (as STAR is a directional measure).  The digitisation used in this method involves measuring the location and direction of each leaf on a tree, as well as using a leaf area meter to measure the size of sampled leaves.

\cite{hadari2004three} investigates the impact of PAR availability and interception on the growth and yield of avocado trees using the "Radiance" lighting simulation software tool.  The approach used is applied on a whole-orchard scale with simplified uniform tree geometries and provided useful conclusions for agricultural practices such as pruning angle and tree height.  However, more accurate geometric modelling would allow higher-resolution simulation and specific recommendations for individual trees.

In this paper, we develop a method for sensing and modelling light interception that is applicable for physical (not just virtual) fruit trees, yet scalable for whole orchards.  By explicitly modelling tree geometry and light conditions, we estimate the distribution of energy throughout the tree in addition to the total light interception.  With the motivation of developing a pruning recommendation system in future work, we here build the underlying model and verify its accuracy compared to ceptometer data.


\section{Method}
\label{sec:method}

\begin{figure}
    \centering
    \includegraphics[width=\linewidth]{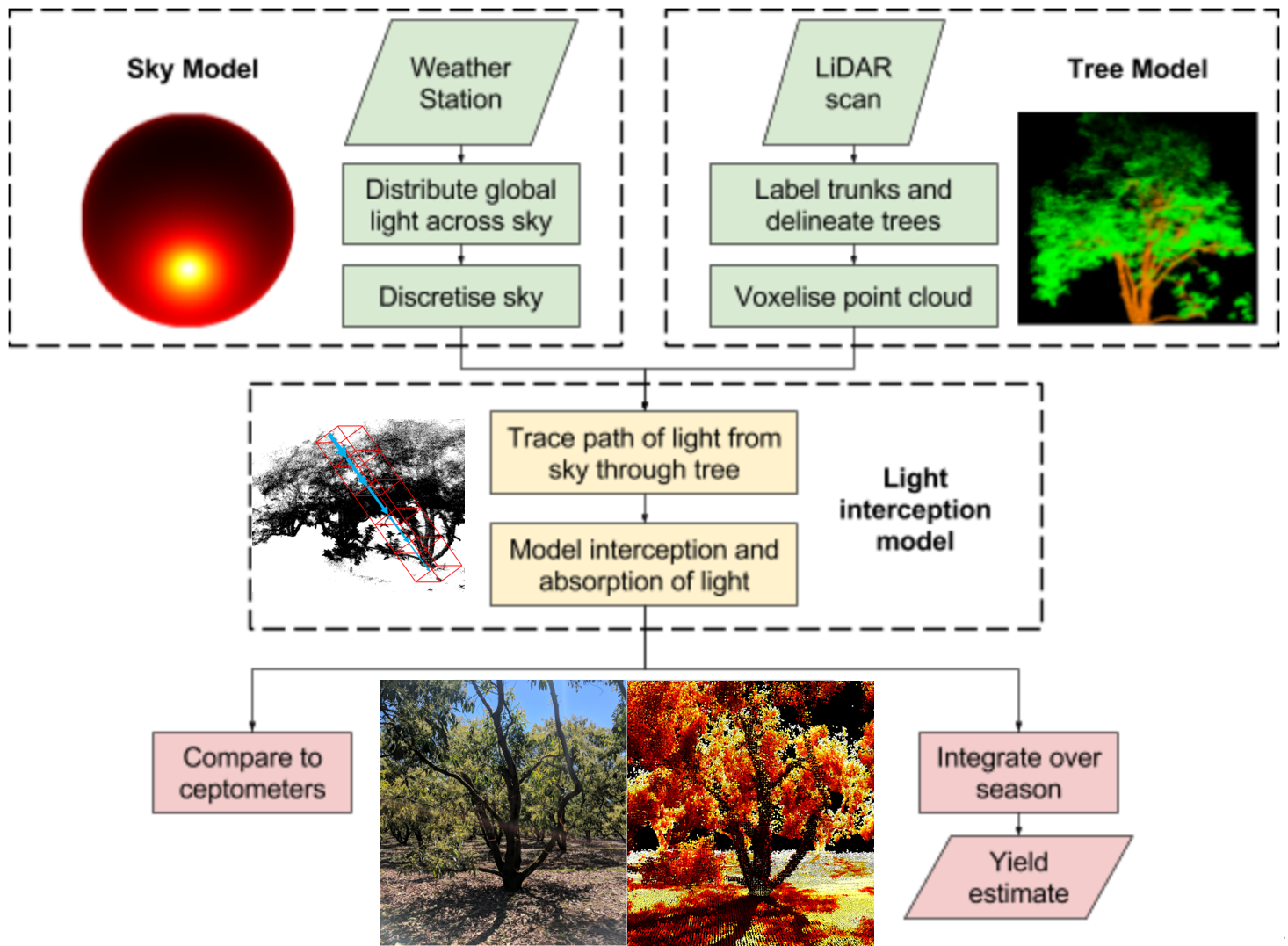}
    \caption{Method used for light interception estimation}
    \label{fig:li-blockdiagram}
\end{figure}

The method used to model light through a given tree canopy is outlined in Figure~\ref{fig:li-blockdiagram}.  On-site weather station data (or national meteorological records) were used to create a model of the sky as a set of discrete light sources with energy values corresponding to a particular time period.  Ray tracing was then used to calculate the distribution and absorption of light in the tree by superimposing the sky model and analysing the path of light from each sky node through the canopy.  Finally, this light model was compared to ground-truth ceptometer data and model parameters were tuned.

\subsection{Data acquisition}
\label{sec:method-data}

Data were gathered from a commercial avocado farm in Bundaberg, Queensland, Australia. Three trees were selected, to include a low, medium and high vigour tree, following the method of \cite{robson2017evaluating}.  These were scanned at multiple times during the growing season in order to capture the changing shape of each tree due to pruning and growth. 

To obtain models of tree geometry, the target trees were scanned using a Zebedee handheld LiDAR (\cite{bosse2012zebedee}).  This scanner consists of a two-dimensional LiDAR scanner which oscillates about the user's hand and uses Simultaneous Localization and Mapping (SLAM) to generate a three-dimensional point cloud scan.  This sensor was selected due to its ability to scan tree geometry throughout the canopy, reducing occlusion.  

Knowledge of the orientation of the tree is critical to simulation of light interception, but the Zebedee LiDAR does not assign a geographical frame of reference to scanned point clouds.  The local scans were therefore aligned to previously obtained georeferenced data from the mobile terrestrial lidar system presented by \cite{stein2016image}.  This combination was chosen to facilitate repeated experiments within a commercial orchard, whereas future work will directly use repeated MTLS scan data.

Weather data for modelling the sky were captured with a Davis Vantage Pro 2 weather station~(\cite{davis2016vantage}) which provided the global irradiance over 30 minute intervals throughout each day.  For periods when the local weather station was offline, we also extracted the daily global solar exposure from the public record provided by the Australian Bureau of Meteorology (\cite{bom2018daily}), and interpolated to relevant times using calibration factors from when both sources were available concurrently.

\begin{figure}
    \centering
    \includegraphics[width=\linewidth]{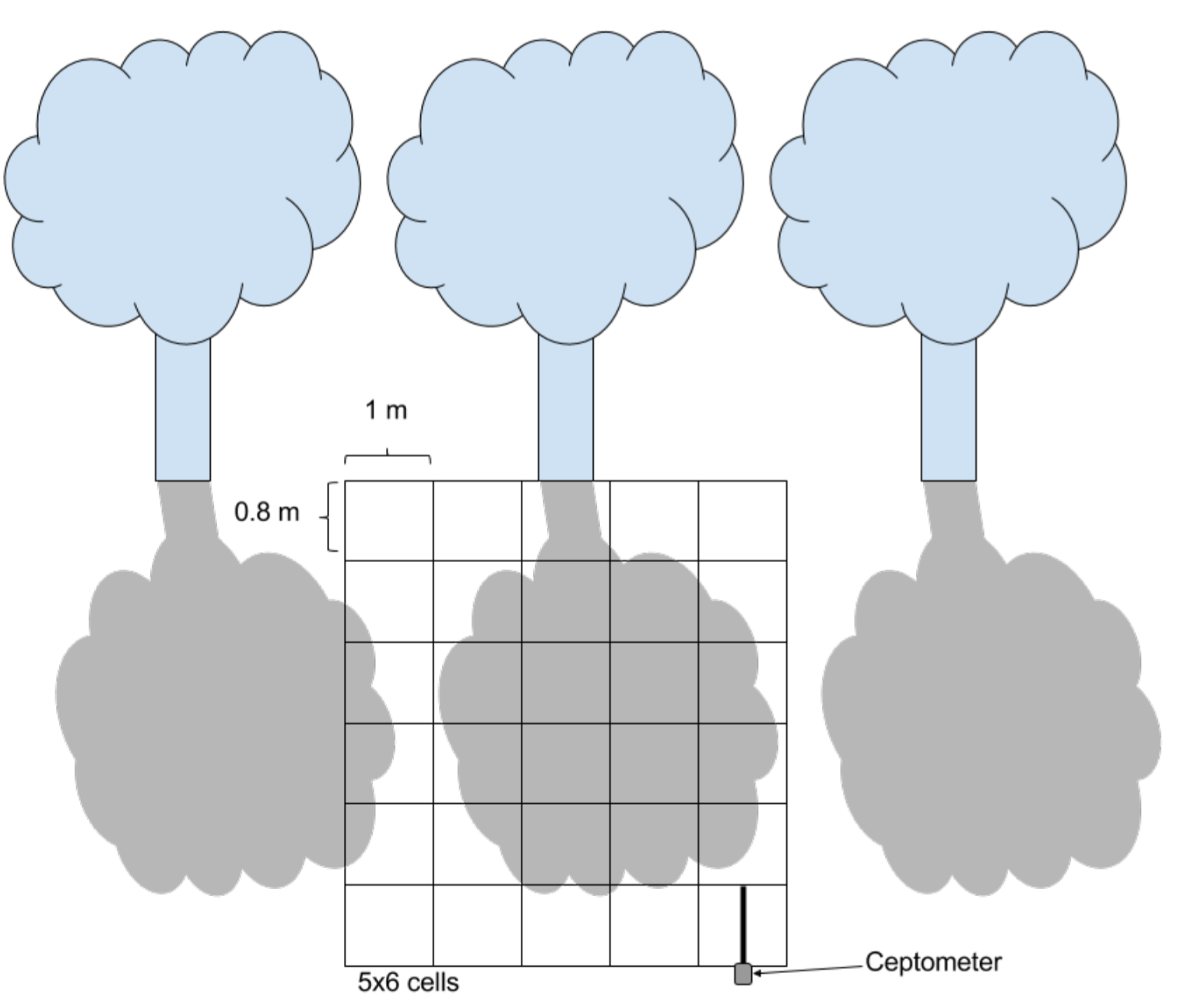}
    \caption{The ceptometer measurements were taken in a regular grid, fixed for each time of day, under specific trees.  Other trees could be close enough to cast a shadow in the grid as well.  Note that this sketch only shows the grid in front of the tree, but measurements were taken in an equivalent grid behind it as well.}
    \label{fig:cepto-grid}
\end{figure}

\begin{figure}
    \centering
    \includegraphics[width=\linewidth,angle=180,origin=c]{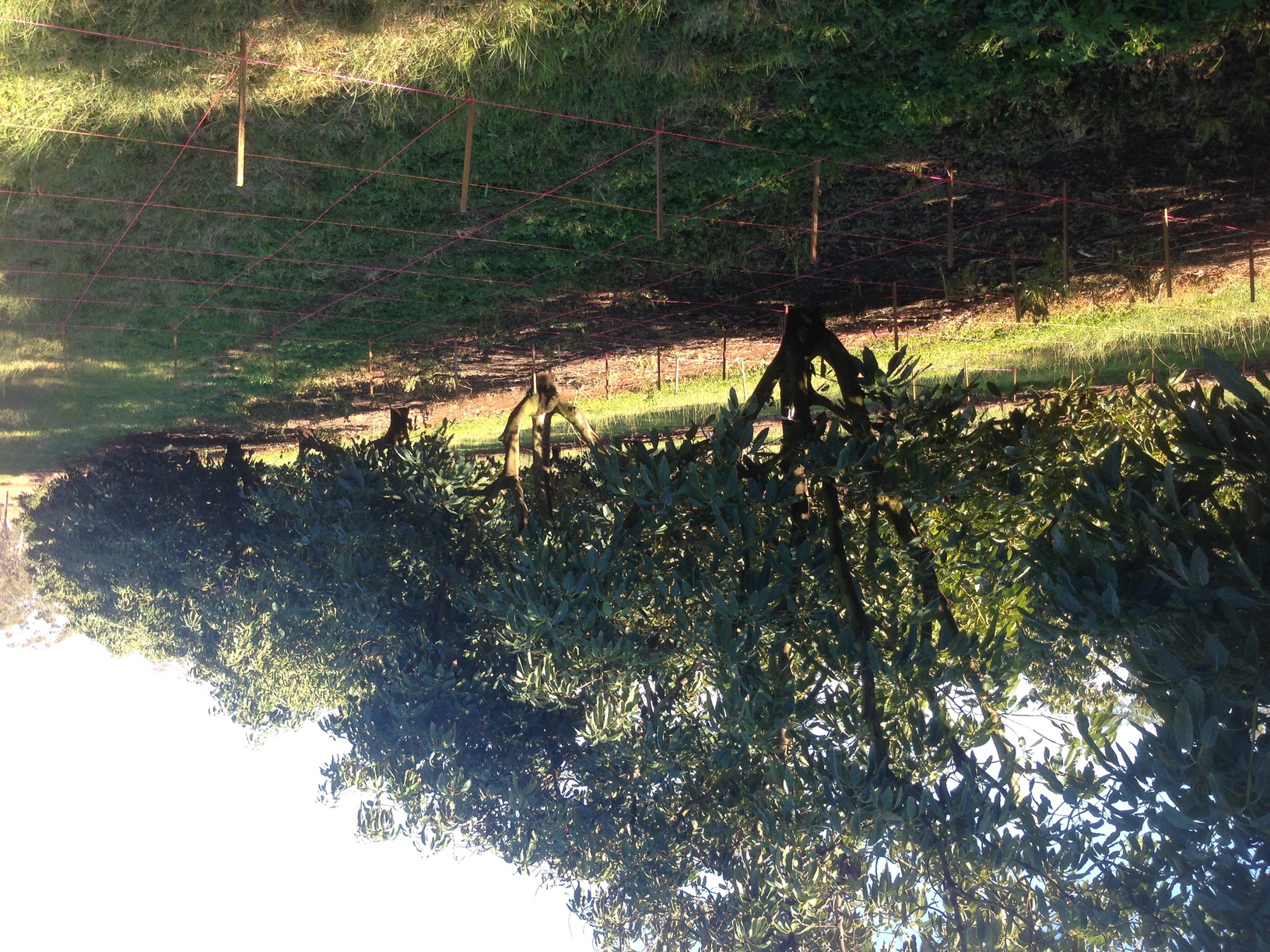}
    \caption{A photograph showing the setup of the ceptometer measurement grid below an avocado tree.}
    \label{fig:cepto-grid-real}
\end{figure}

\begin{figure}
    \centering
    \includegraphics[width=\linewidth]{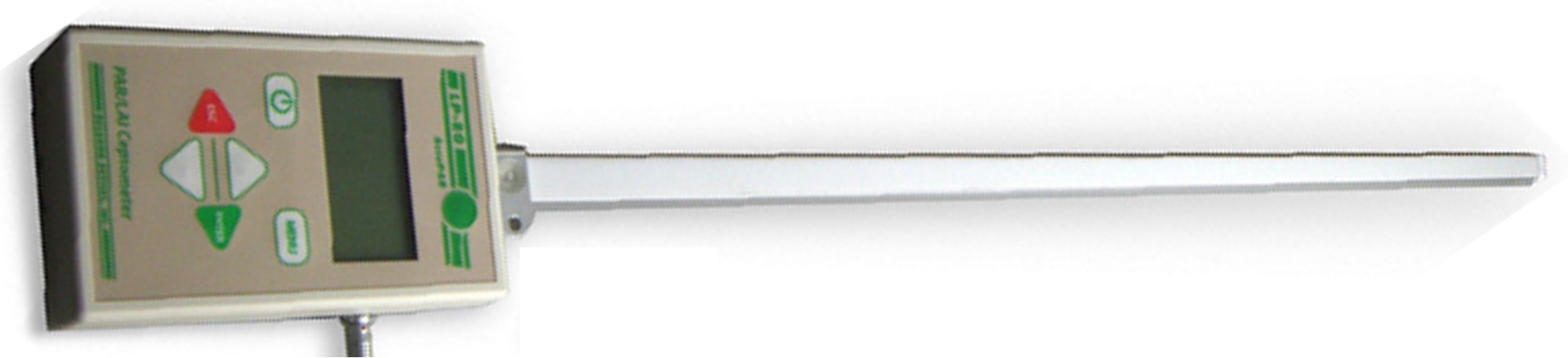}
    \caption{The ceptometer used to gather data.  Eight light sensors measuring PAR in $\mu$ mol s\textsuperscript{-1} m\textsuperscript{-2} are evenly spaced along its shaft.}
    \label{fig:ceptometer}
\end{figure}

To validate the model, ceptometer measurements were taken in a regular grid below the data trees, with a 1m row space along the orchard row and 0.8m spacing perpendicular to the row direction, as shown in Figures \ref{fig:cepto-grid} and \ref{fig:cepto-grid-real}.  Measurements were taken with the 80cm long ceptometer (see Figure~\ref{fig:ceptometer}) perpendicular to the tree row.  The ceptometer sensor takes eight readings evenly spaced along its length, measuring PAR in $\mu$ mol s\textsuperscript{-1} m\textsuperscript{-2}.  In addition to measurements within the grid, a second identical ceptometer was placed at the edge of the orchard block, in a constantly unshaded area, logging data continuously at one-minute intervals.

This process was performed on the target trees multiple times a day during four different days in 2016-17.  Overall, 33 distinct datasets were available.  For 15 of these, data were available for each of the eight sensors in the ceptometer, and for the other 18 only averages of the eight were available, so in our comparisons we use the average of all eight readings to form each data point.  Open-air ceptometer measurements were available for 24 datasets, and Vantage Pro weather station data were available for all but 9 datasets, for which Bureau of Meteorology data were used.

The method used here for LiDAR scanning captured the data trees in great detail, but not the neighbouring trees.  As such, the ceptometer data from the south side of the tree were representative of the light interception from the data tree but the north side was shaded by geometries that have not been modelled.  Therefore north side ceptometer measurements were excluded from our experiments.

\subsection{Sky modelling}
\label{sec:method-sky-model}

The sky was modelled as a set of discrete light sources evenly spaced on a hemispherical surface, which was generated deterministically using the method presented by \cite{weisstein2003geodesic} to produce a geodesic sphere.  The sky resolution was controlled by a model parameter $S$, representing the number of discrete points in the sky.


Global irradiance at a particular point on Earth, aquired as described in Section~\ref{sec:method-data}, can be decomposed into a direct component which measures the light travelling straight from the sun, and a diffuse component which integrates light from all indirect pathways travelled between the sun and that point.  The diffuse and direct components were calculated using a method presented by \cite{ridley2010modelling}:

\begin{equation}
    D_{frac}=\frac{1}{1+e^{-5.38+6.63k_t+0.006AST-0.007\alpha+1.75K_t+1.31\phi}}
    \label{eq:dfrac}
\end{equation}
\begin{equation}
    k_t = \frac{I_{global}}{H_0}
\end{equation}
\begin{equation}
    K_t = \frac{\sum_{h=1}^{24} I_{global}}{\sum_{h=1}^{24} H_0}
\end{equation}
\begin{equation}
    \phi = \frac{k_{t-1}+k_{t+1}}{2},
\end{equation}

where $D_{frac}$ is the diffuse fraction of light (which can be used to calculate the absolute diffuse and direct irradiance), $k_t$ clearness index, $I_{global}$ global irradiance on Earth, $H_0$ extraterrestrial irradiance, $K_t$ daily clearness, $\phi$ persistence factor, $\alpha$ the solar angle and $AST$ the apparent solar time.

According to \cite{kennewell2016equation} $AST$ can be computed from mean solar time (UTC time):

\begin{equation}
    B = \frac{360(N-81)}{365} [degrees] 
\end{equation}
\begin{equation}
    AST = UTC + 9.87\sin(2B) - 7.67\sin(B + 78.7),
\end{equation}

where $N$ is day of year counting from January 1st when $N = 1$

The extraterrestrial irradiance $H_0$, which is the irradiance at the entry point of the atmosphere, can be estimated from the solar constant $S_C \approx 1370W/m^2$ and taking the deviation from the mean distance between the sun and the Earth on day $N$ of the year into account ($N \in [1 : 366]$ counting from January 1st and considering leap years).  This gives the following expression:

\begin{equation}
    H_0 = S_C(1+0.033412\cos(2\pi(N-3)/365)).
\end{equation}

\begin{figure}
    \centering
  \subfloat[Clear day]{%
       \includegraphics[width=0.8\linewidth]{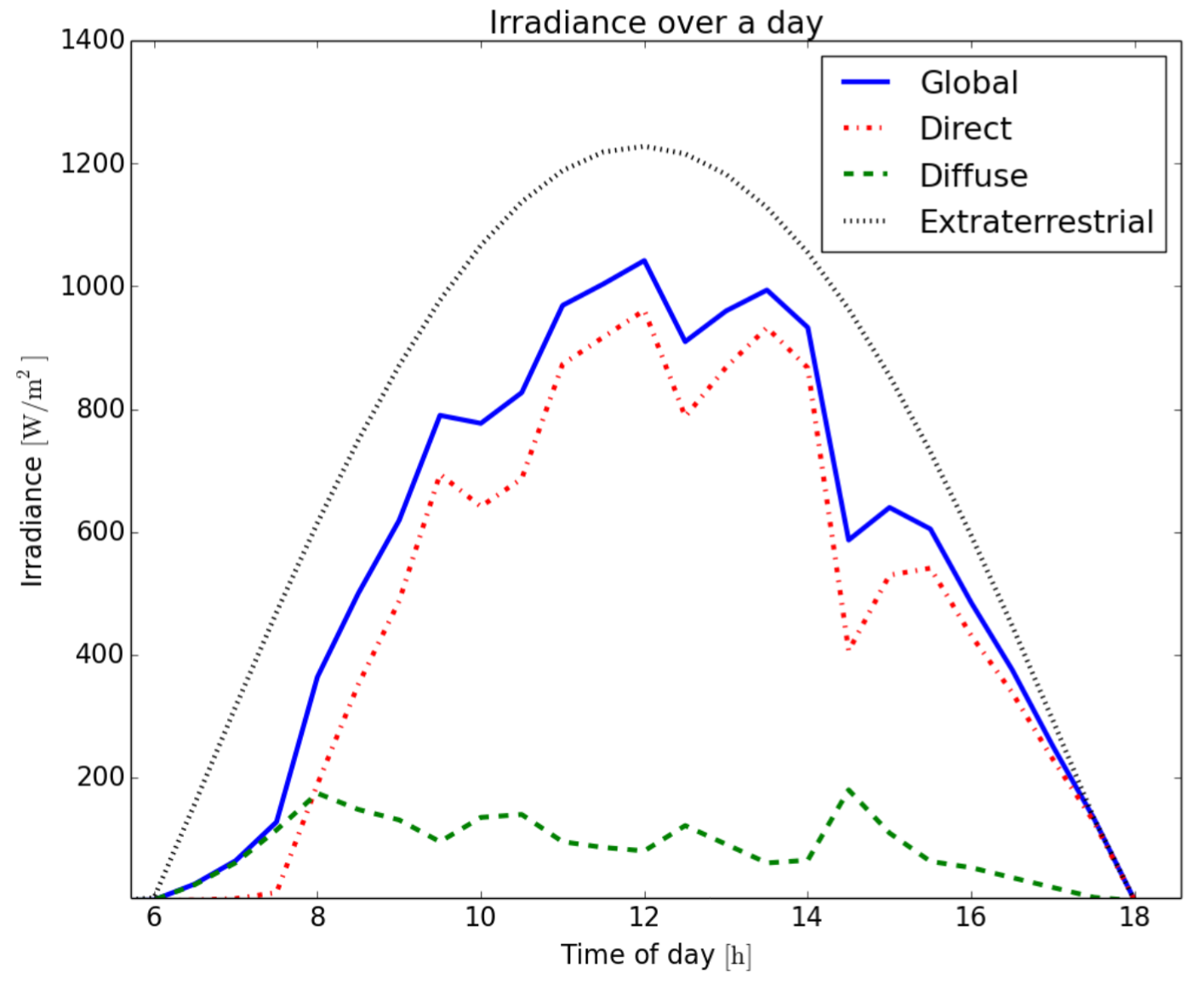}
    \label{fig:irradiance-clear}}\hfill
  \subfloat[Cloudy day]{%
        \includegraphics[width=0.8\linewidth]{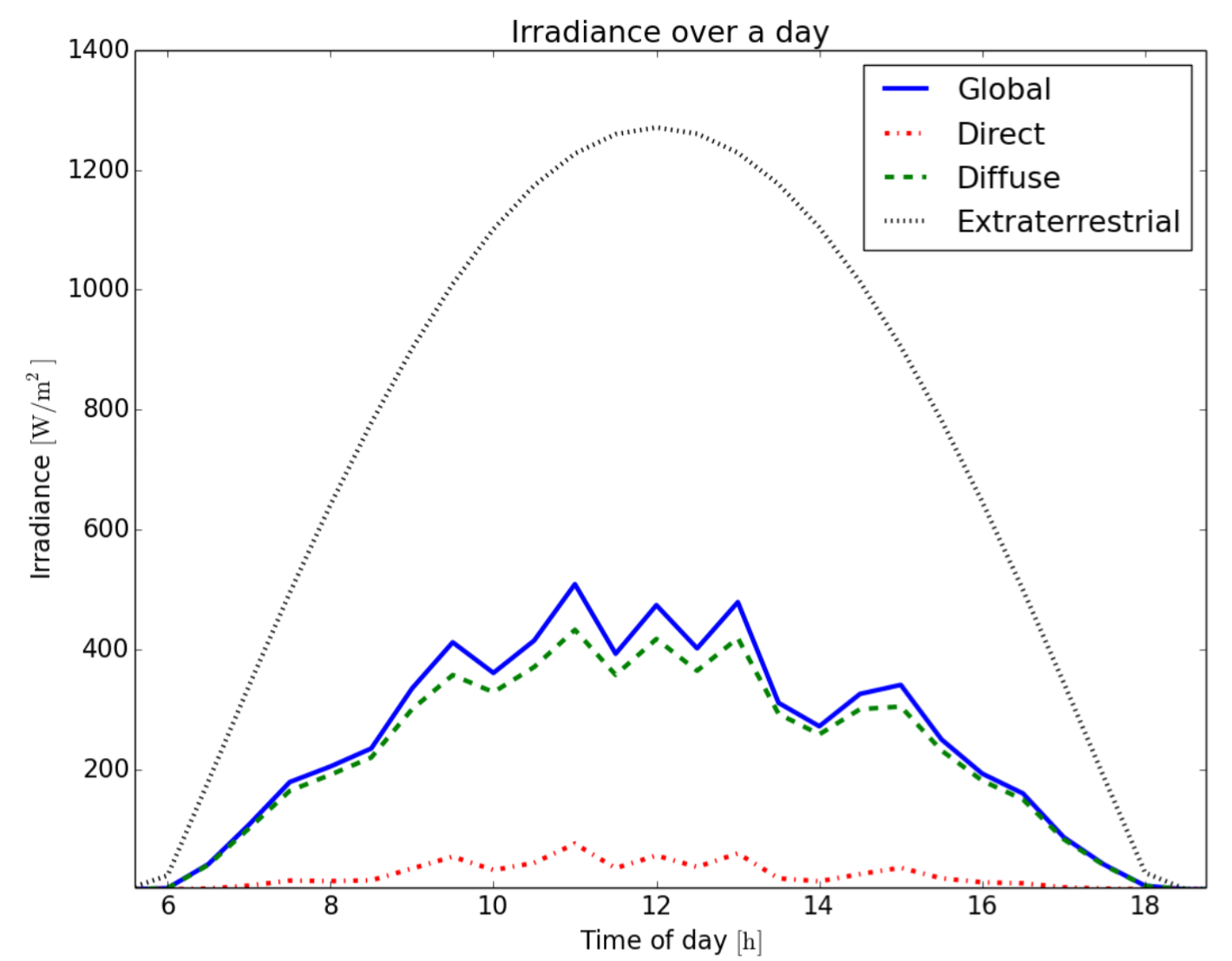}
    \label{fig:irradiance-cloudy}}\hfill
  \subfloat[Mixed day]{%
        \includegraphics[width=0.8\linewidth]{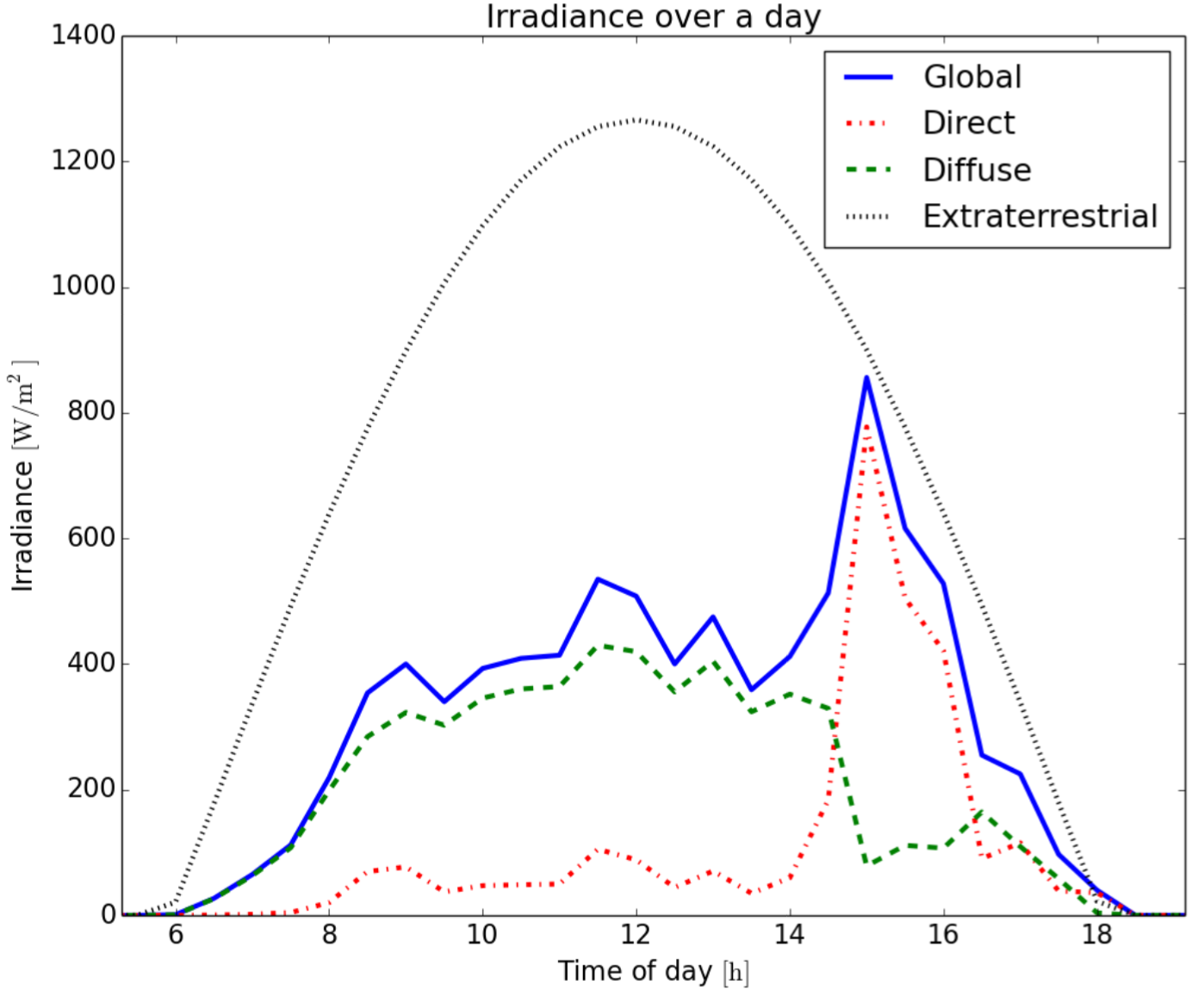}
    \label{fig:irradiance-mixed}}\hfill
  \caption{Irradiances for different sky types.  The global measurements are from the weather station at the farm, the diffuse and direct components are estimated.}
  \label{fig:global-irradiance} 
\end{figure}

Figure \ref{fig:global-irradiance} demonstrates how global irradiance was split into direct and diffuse components for different days using the method described above.

For any latitude and longitude on Earth, it is trivial to calculate the current solar position as described by \cite{reda2004solar}.  In this work, the Python package Pysolar (\cite{stafford2014pysolar}) was used to obtain the solar elevation and azimuth.  This allowed us to create a new sky node at this position, to which the direct irradiance component was added.  

The diffuse component can be represented as $S$ separate light sources placed at the vertices of the discretised sky, using a diffuse sky distribution model.  The CIE General Sky Standard, specified by \cite{darula2002cie}, states how to estimate the relative diffuse luminance between two different points in the sky, here denoted $L_{rel}$:

\begin{equation}
    L_{rel} = f(\chi)\phi(Z)
    \label{eq:lrel}
\end{equation}

\begin{equation}
    f(\chi) = 1 + c(\exp(d\chi) - \exp(d\pi/2)) + e\cos^2(\chi)
    \label{eq:cde}
\end{equation}

\begin{equation}
    \phi(Z) = 1 + a\exp(b/\cos(Z))
    \label{eq:ab}
\end{equation}

\begin{equation}
    \chi = \arccos(\cos(Z_s)\cos(Z) + \sin(Z_s)\sin(Z)\cos(Az)),
\end{equation}

where $L_{rel}$ is the relative luminance of a certain sky location, $f(\chi)$ is the scattering indicatrix, $\phi(Z)$ is the luminance gradation and $\chi$ is the great arc distance between the sun and the sky element of interest.  The inputs required are: sun zenith angle $Z_s$, element of interest's zenith angle $Z$, difference in azimuthal angle between the sun and the element of interest $Az$ and five parameters $a,b,c,d,e$ describing the type of sky (overcast, clear, polluted, etc).

Which sky model to use (parameters $a, b, c, d, e$ in \eqref{eq:cde} and \eqref{eq:ab}) was determined from the diffuse fraction of light $D_{frac}$ using the rules shown in Table~\ref{tab:cie-types} and use a subset of the 12 CIE sky types.

\begin{table}[h]
    \centering
    \begin{tabular}{cccccccc}
         $D_{frac}$ & Type & Description & $a$ & $b$ & $c$ & $d$ & $e$ \\
         \hline
         $[0.00,0.25]$ & 12 & Standard clear &  -1 & -0.32 & 10 & -3 & 0.45 \\
         $(0.25,0.50]$ & 11 & White-blue sky & -1 & -0.55 & 10 & -3 & 0.45 \\
         $(0.50,0.75]$ & 7 & Partly cloudy & 0 & -1.0 & 5 & -2.5 & 0.30 \\
         $(0.75,1.00]$ & 1 & Standard overcast & 4 & -0.7 & 2 & -1.5 & 0.15 \\
    \end{tabular}
    \caption{CIE parameters used by the value of $D_{frac}$}
    \label{tab:cie-types}
\end{table}

As $L_{rel}$ in \eqref{eq:lrel} is the relative luminance, a reference measurement is needed to obtain the actual luminance. \cite{darula2002cie} normalises $L_{rel}$ by the luminance at the zenith $L_{zenith}$:

\begin{equation}
    L_{zenith} = f(Z_s)\phi(0),
\end{equation}

which means that the actual luminance can be obtained by normalising each relative luminance by a reference measurement at zenith.

The model presented by \cite{darula2002cie} is intended to distribute a single central luminance measure across the sky, while we instead had the integral of diffuse irradiance over the whole sky.  Since the sky node area is uniform and the spectrum of the light source is assumed to be constant, irradiance and luminance differed only by a constant factor and we could use the same equations for irradiance.  We distributed the integral through the nodes using the relative equation and an integrated normalisation factor (since no zenith measurement was available) using:

\begin{equation}
    I_{diffuse,node} = \frac{I_{diffuse}}{\sum_{sky} L_{rel}} L_{rel,node}.
\end{equation}

\begin{figure}
    \centering
  \subfloat[Clear sky]{%
       \includegraphics[width=0.4\linewidth]{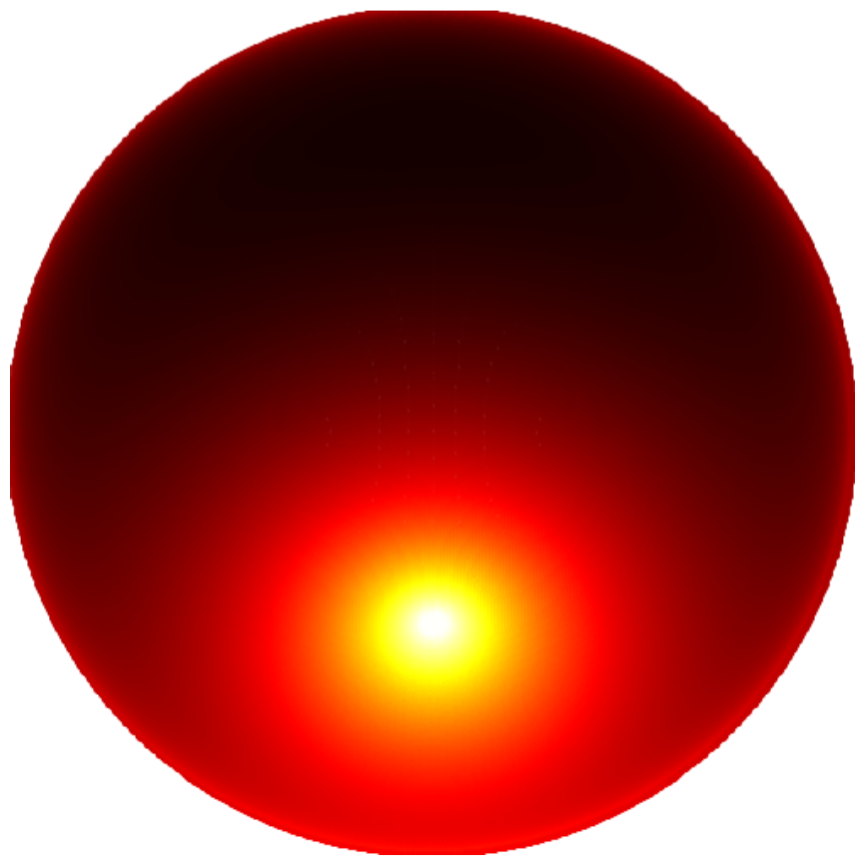}
		\label{fig:modelled-sky-clear}\hfill}
  \subfloat[Overcast sky]{%
        \includegraphics[width=0.4\linewidth]{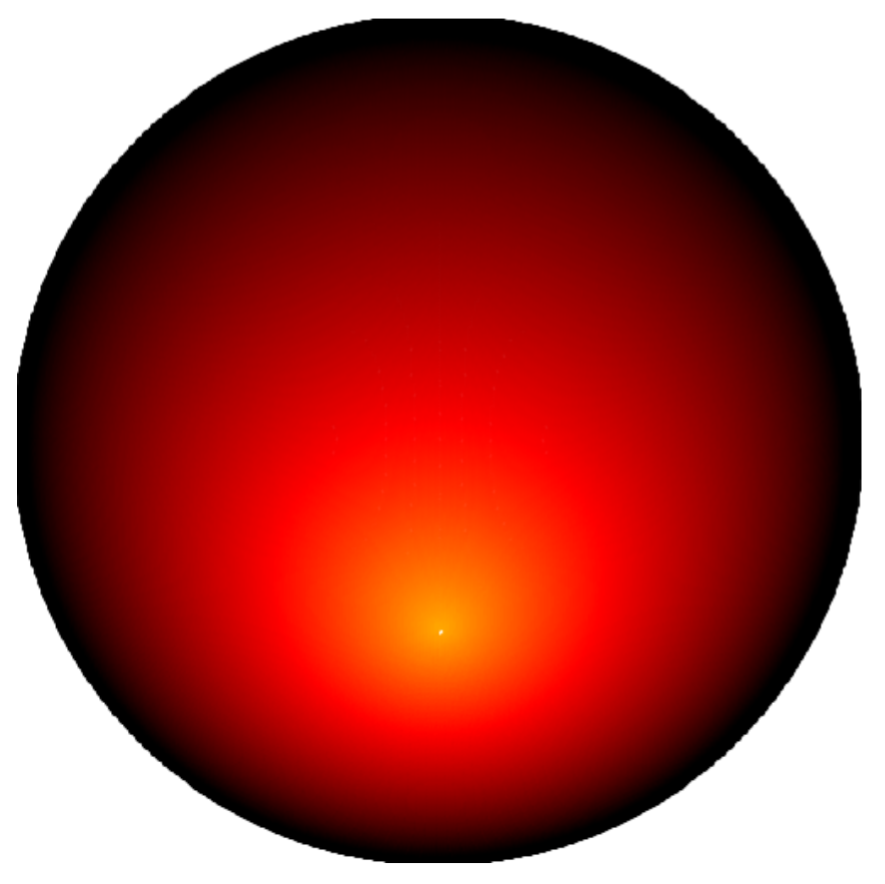}
        \label{fig:modelled-sky-cloudy}}
  \subfloat{%
        \includegraphics[width=0.125\linewidth]{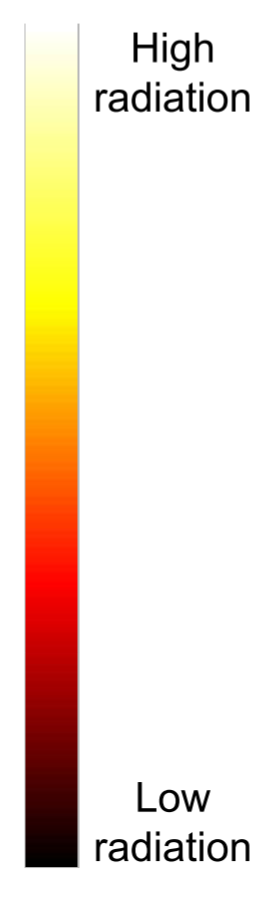}
        \label{fig:modelled-sky-mixed}}
  \caption{Distribution of diffuse light over a discretised clear and overcast sky at 14:00 on a winter day.  The colour scale in both sub figures represent the same range of radiation.  CIE standard skies 12 and 1 from~\cite{darula2002cie} respectively were used to generate the distribution.}
  \label{fig:modelled-sky} 
\end{figure}

Figure~\ref{fig:modelled-sky} shows a visualisation of the diffuse light distribution over a sky on a clear and overcast day generated by this method.  Both visualisations are from the same day and time, and hence have the same total amount of diffuse light, and were generated using the same colour scale.  The clear sky demonstrated that there is a higher concentration of diffuse light close to the sun, as opposed to the overcast sky, where the light is distributed over a larger area around the position of the sun.

\begin{figure}
    \centering
  \subfloat[Single day]{%
       \includegraphics[width=0.4\linewidth]{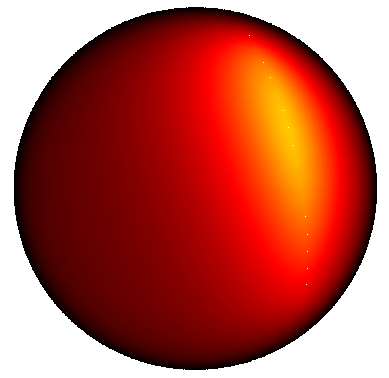}
		\label{fig:modelled-sky-day}\hfill}
  \subfloat[Full year]{%
        \includegraphics[width=0.4\linewidth]{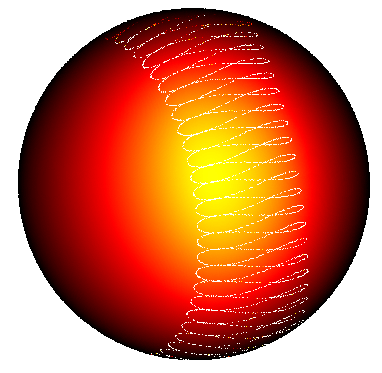}
        \label{fig:modelled-sky-season}}
  \subfloat{%
        \includegraphics[width=0.125\linewidth]{figures/modelled-sky-scale}}
  \caption{Two composite skies generated by summing sky models at discrete time points during a single day and a full year.  Note the image was scaled so that the pattern of the diffuse sky is visible, since the direct component of light is significantly brighter.  The magnitude of radiation in each figure is also different due to the timescale involved.  Best viewed digitally with zooming.}
    \label{fig:composite-sky}
\end{figure}

Whenever a temporal integration is required, for instance to calculate the total light over a growing season, a composite sky for a given time span can be computed, where we model the total energy in the sky over a known interval rather than the instantaneous power.  This composite is simply the sum of skies generated at regular time intervals, and an example of this can be seen in Figure~\ref{fig:composite-sky}.  For composite skies, adding a sun node for the exact solar position as done for instantaneous simulation would generate an infeasibly high-resolution sky.  Thus we instead added the direct component of light to the nearest existing sky node when integrating over multiple intervals.


\subsection{Point cloud processing}
\label{sec:method-pointcloud}

As described in Section~\ref{sec:method-data}, the geometry of the trees on the farm was represented by geo-referenced LiDAR data captured by a handheld Zebedee scanner.  Non-uniform point cloud sampling densities arose from the variable range from sensor to target, and the complex patterns of foreground occlusions. Consequently, the data density was regularised by voxelisation as described by \cite{douillard2011segmentation}, with each voxel represented as a cube with side length $s_{vox}$.  As a form of noise rejection, a model parameter $w_{vox}$ was introduced as the minimum number of points within a voxel for it to be included.  We also maintained a correspondence between each original LiDAR point and its associated voxel, allowing computationally expensive operations such as ray tracing to be performed in the smaller voxel space and later redistributed to the original data with no loss of resolution.

\begin{figure}
    \centering
    \includegraphics[width=\linewidth]{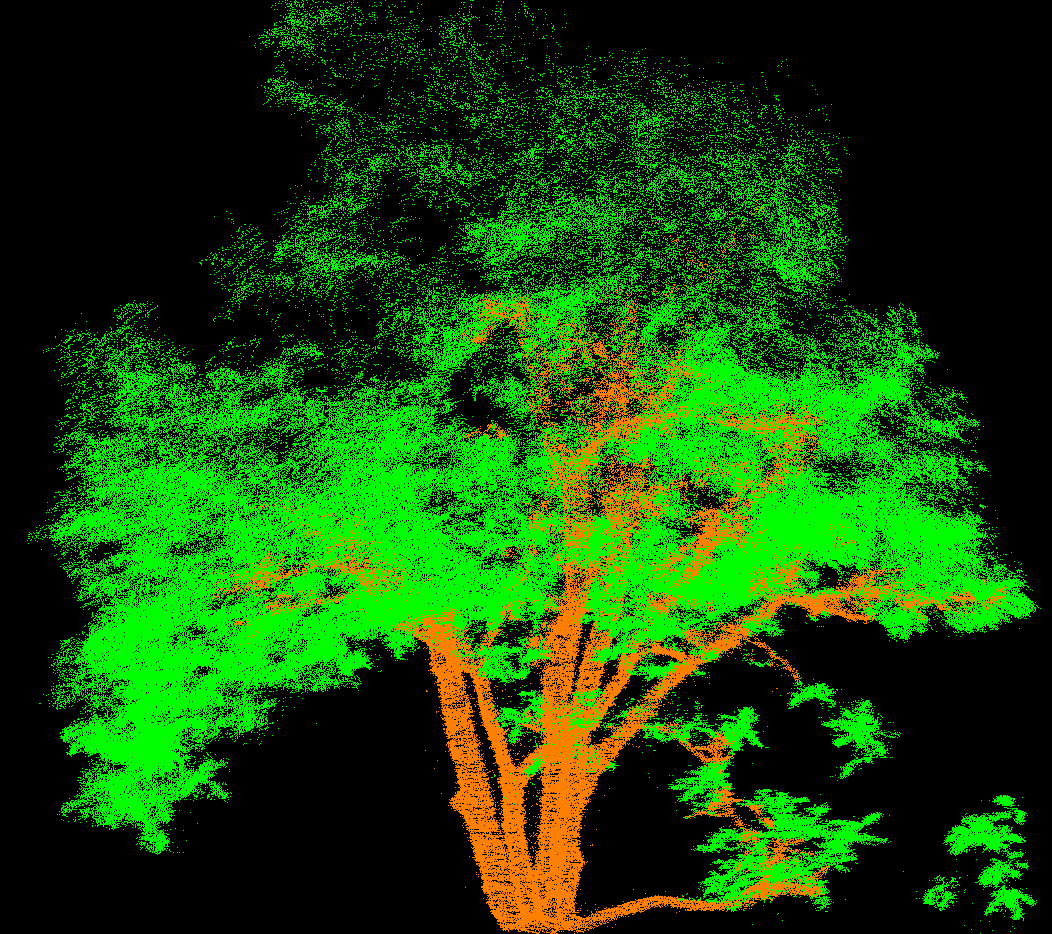}
    \caption{A point cloud model of a target tree with points manually labelled as woody matter and foliage.}
    \label{fig:trunk-label}
\end{figure}

The point clouds were manually segmented into branches and foliage as shown in Figure~\ref{fig:trunk-label}, allowing different parameters to be applied for each\footnote{Future work will consider automated approaches such as those presented by \cite{Lalonde2006}, \cite{Olofsson2014} and \cite{Ma2016}}.  These parameters were the transmission coefficient $\beta$ and the absorption coefficient $\alpha$, which estimate the proportion of light which passed through and was absorbed by the voxel respectively.  The segmentation reflects that branches are opaque ($\beta_{b} = 0$) and not photosynthetically active ($\alpha_{b} = 0$), as distinct from regions of foliage which have non-zero values that sum to one ($\beta_{f} > 0$, $\alpha_{f} + \beta_{f} = 1$).

\subsection{Interception of radiation by a tree}
\label{sec:method-raytracing}

\begin{figure}
    \centering
    \includegraphics[width=\linewidth]{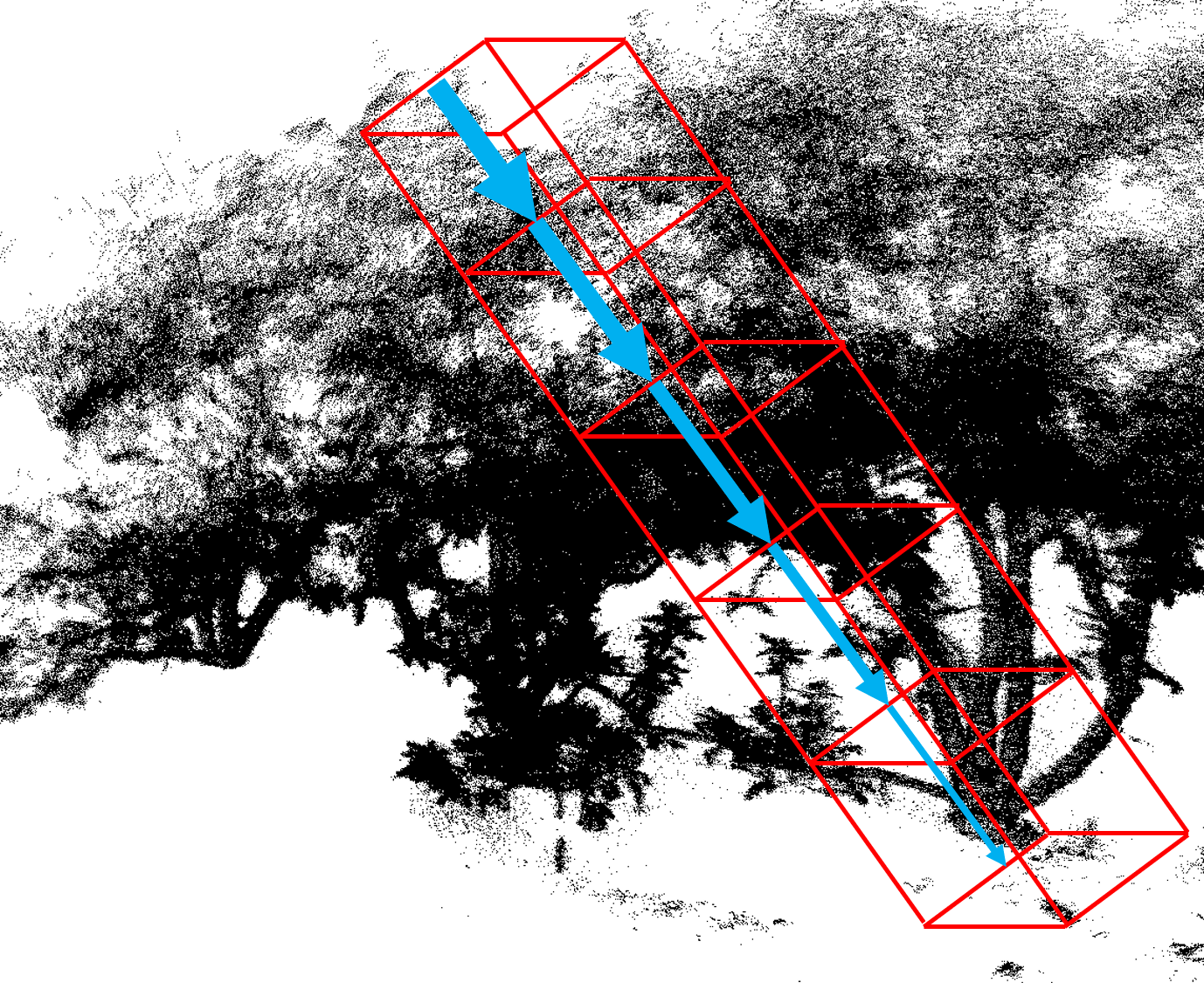}
    \caption{Illustration of the raytracing process.  The red wireframe represents a single column of voxels through the point cloud.  The blue arrows represent the available irradiance diminishing as it travels through the voxels.  Smaller voxels were typically used when processing the data.}
    \label{fig:voxels}
\end{figure}

Knowing the energy in each sky node and assuming parallel rays, we traced the path of light through the tree from each node individually. The point cloud was first voxelised with a grid that was oriented towards the node, and voxels were organised into columns parallel to the light ray which were traversed sequentially to calculate the energy available to each voxel.  This process is illustrated in Figure~\ref{fig:voxels}.

The quantum of energy in each sky node was distributed to the voxels along each sky-to-ground path, using a method based on that suggested by  \cite{charles1982physiological}:

\begin{equation}
\label{eq:charles}
    I = I_0(s + (1-s)\beta)^{n-1},
\end{equation}

which estimates the downward light flux density falling on the $n$th voxel in a column knowing the incident irradiance $I_0$ and assuming a constant transmission coefficient $\beta$.  In our work, the gap fraction $s$ in \eqref{eq:charles} was ignored because it is explicitly represented by the voxelised geometry of the tree.  In other words, occupied voxels were considered to have a gap fraction of 1 and the gap fraction of larger regions is represented as the ratio of occupied to unoccupied voxels.  Furthermore, we did not assume a constant transmission coefficient as each voxel had a unique coefficient $\beta_i$ calculated as the average coefficient of its constituent points.  Thus we calculated the irradiance available to the $n$th voxel in a column as:

\begin{equation}
    \label{eq:incident-intensity-complex}
    I_n = I_0\prod_{i=0}^{n-1} \beta_i.
\end{equation}

The irradiance absorbed $I_{abs}$ by this voxel can then be derived from Equation \eqref{eq:incident-intensity-complex} as $I_{abs,n} = \alpha_n I_n$.
For comparison to ground-truth sensor data, we used the instantaneous irradiance absorbed ($W m^{-2}$) in this form.  However, the absolute energy ($J$) is required for integration over a growing season, and this can easily be found as $E_{n}$ using:

\begin{equation}
    E_{n} = I_{abs,n}A_{vox}\Delta t,
\end{equation}

where $A_{vox}$ is the voxel area facing the sky node, calculated as the voxel side length squared, and $\Delta t$ is the time step which the incident irradiance is valid for.

The mapping of lidar points to voxels changes every time the point cloud is oriented to a different sky node. Consequently it is not possible to tally the energy across all sky nodes in a consistent way per voxel. Instead energy values were stored in the original lidar points within the voxel.  The voxel energy was distributed evenly to all points, while the irradiance added to each point was equal to the total irradiance of the voxel.  In this way, the total light available became the sum of the contribution from every sky node, and we calculated total energy by summing all lidar points or sample irradiance at any point.

\subsection{Validation by Comparison to Ceptometer Data}

\begin{figure}
	\centering
	\includegraphics[width=\linewidth]{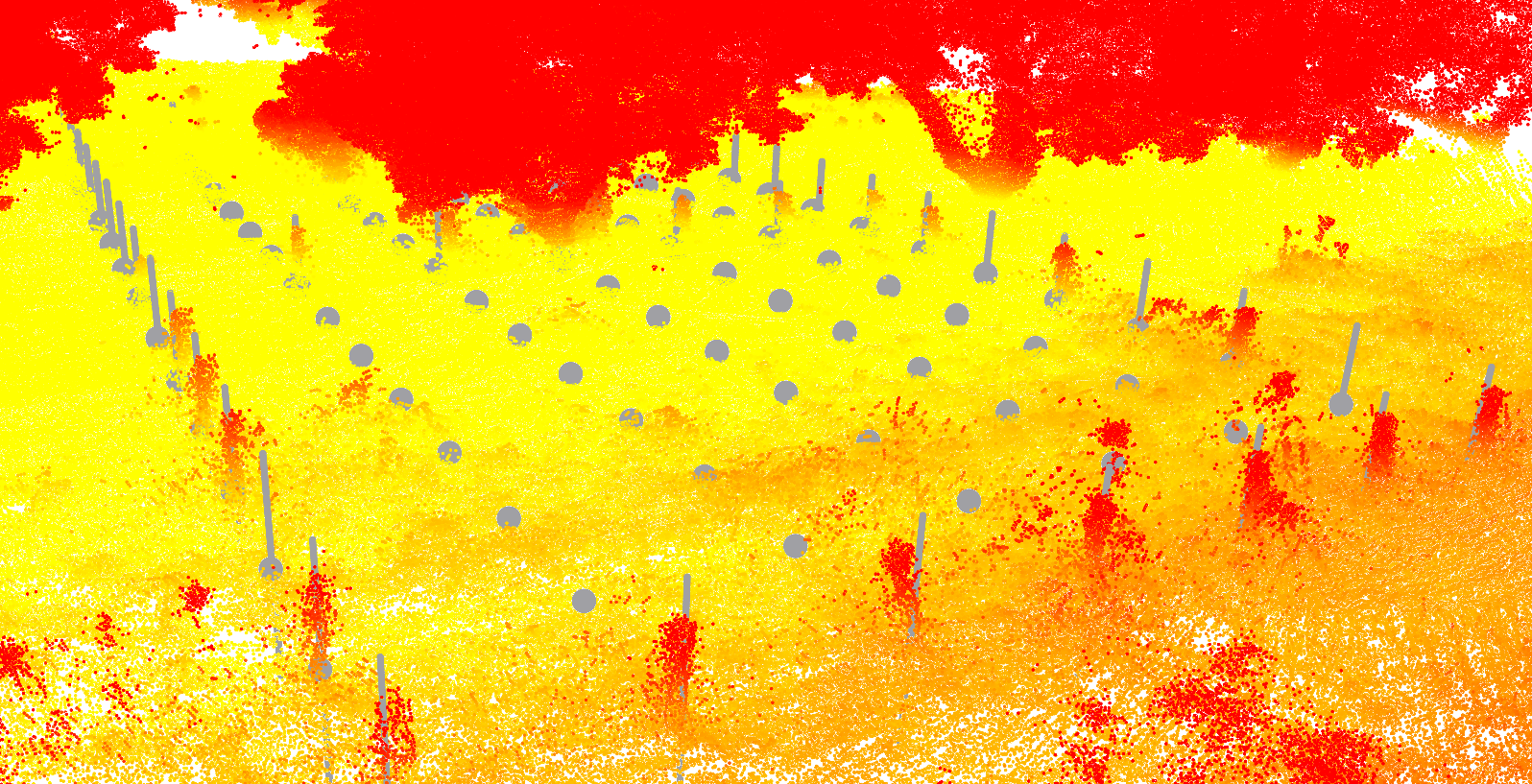}
	\caption{Ceptometer locations overlaid on LiDAR point cloud. Initial offset achieved by aligning the corner stakes and refined by optimisation.}
	\label{fig:cepto-grid-pc}
\end{figure}

In order to validate the model, we compared estimated energy values at known points with ceptometer data captured within a staked-out grid as described in Section~\ref{sec:method-data}.  Ceptometer locations within the point cloud were interpolated from the positions of the physical stakes visible in the LiDAR scans as shown in Figure~\ref{fig:cepto-grid-pc}.  The irradiances stored within the point cloud in the immediate neighbourhood of the ceptometer locations were then averaged to obtain a reading which can be compared to the ground truth values.

Conversion from full-spectrum irradiance ($W m^{-2}$) to the Photosynthetically Active Radiation (PAR) range ($\mu$ mol $s^{-1}$ $m^{-2}$) measured by the ceptometers involves scaling to convert to mols and then taking a subsection of the full spectrum.  These operations can be combined by multiplication of a constant factor, assuming that light has a constant spectrum.  According to \cite{murphy2011maximum} and \cite{thimijan1983photometric}, conversion from irradiance to PAR can be done as:

\begin{equation}
\label{eq:parconvert}
    I_{PAR} = 1.72 I_{irr},
\end{equation}
where $I_{irr}$ is irradiance in $W m^{-2}$ and $I_{PAR}$ is PAR in$\mu$ mol $s^{-1}$ $m^{-2}$.

As described in Section~\ref{sec:method-data}, a second ceptometer provided a measurement of open air PAR at one-minute intervals for the duration of experimentation.  We used data from this ceptometer for experimental validation against sub-canopy ceptometer data taken at specific instances in time.  This was necessary to compensate for local, short-term, micro-climactic variations such as patchy cloud cover that can vary more rapidly and locally than the weather station can compensate for.  Comparing this open air measurement of PAR, denoted $O_{PAR}$, and integrating the sky model to calculate an estimated open air irradiance value $O_{irr}$, we calculated $I_{PAR}$ using:

\begin{equation}
    I_{PAR} = I_{irr} (O_{PAR}/O_{irr}).
\end{equation}

This was necessary for our experimental validation against instantaneous data, whereas for whole season calculations when continuous open air ceptometer data was unavailable, Equation \eqref{eq:parconvert} was used instead, since short-term phenomena are averaged out over whole days, months and seasons.

The model produced estimates in PAR for the ceptometer measurement at a particular grid location with a given tree, date and time, we directly compared these with the actual measurements in a scatter plot, and calculated a best-fit line.  The coefficient of determination ($R^2$) and gradient ($m$) were calculated and compared to ideal values (1.0 in both cases).  We also calculated the root mean squared error (RMSE) of all data points to the best-fit line.

\subsection{Parameter Optimisation}

The method introduced in Section~\ref{sec:method} has a number of free parameters, which are summarised in Table~\ref{tab:method-tuning}.  These parameters were optimised to maximise $R^2$ of the comparison to ceptometer measurements.

The ceptometers used to generate the ground-truth data were aligned manually to the LiDAR scan by identifying the stakes in the corner of the ceptometer grid within the lidar data.  Any error in placement of the physical or virtual stakes propagates to the query points used in simulation, so we varied the alignment by different x and y offsets to maximise $R^2$ in comparison to ceptometer data.

\begin{table}[h]
    \centering
    \begin{tabular}{cc}
        Parameter & Name \\
        \hline 
         $S$ & Sky resolution \\
         $\beta_{f}$ & Transmission coefficient for leafy matter \\
         $s_{vox}$ & Voxel size \\
         $w_{vox}$ & Minimum voxel weight \\
         $[\Delta x,\Delta y]$ & Virtual ceptometer offset
    \end{tabular}
    \caption{Model parameters which require tuning}
    \label{tab:method-tuning}
\end{table}

Further experimentation was undertaken to validate different sub-components of the model output, to justify the added complexity of each component.  The primary validation tests involved purposefully changing the major sub-components of the model to test the sensitivity of the output, and are summarised in Table~\ref{tab:method-validation}.

\begin{table}[h]
    \centering
    \begin{tabular}{p{0.4\linewidth}p{0.4\linewidth}}
         Experiment & Hypothesis \\
         \hline
         Point cloud of incorrect tree & Inter-tree differences affect results \\
         Incorrect time of day & Changes in the sky model during the day affect the results \\
         Incorrect date & Differences in sky on different days affect the results\\
         Rotated trunk (various degrees) & Intra-tree differences affect results \\
         Dedicated sun node & Exact sun position affects results \\
		No diffuse light & Splitting total light into diffuse and direct components affects results
    \end{tabular}
    \caption{Validation experiments, and the hypotheses they are designed to test}
    \label{tab:method-validation}
\end{table}

One of these sub-components is the specific LiDAR scan used.  By altering this, we measured the importance of inter-tree geometric differences and validated the conclusion that all avocado trees are not so similar that the specific tree geometry is unimportant. Furthermore, we believe modelling the structures of any one tree must be done accurately to achieve accurate light distribution results.  To validate this, we rotated the virtual tree within its environment.   For example, this tests the alternative hypothesis that lidar scans are merely capturing approximate geometry such as tree height, and that internal structural details are unimportant.

The time of day and date were also varied to investigate the effect of how the sky changes both during a single day and during a season.  We ascertained the importance of inserting an additional sky node at the exact solar position by comparing results when direct irradiance was instead snapped to the nearest sky node.  Finally, if the critical quantity for calculating results is the direct light from the solar sky node, the light from diffuse sources may not be required for results.  To test this, we performed an experiment where the sky contained no diffuse nodes, rather all total light was accumulated in the solar node.


\section{Results}
\label{sec:results}

\begin{figure}
    \centering
    \subfloat[1 ceptometer per data point]{%
    \includegraphics[width=\linewidth]{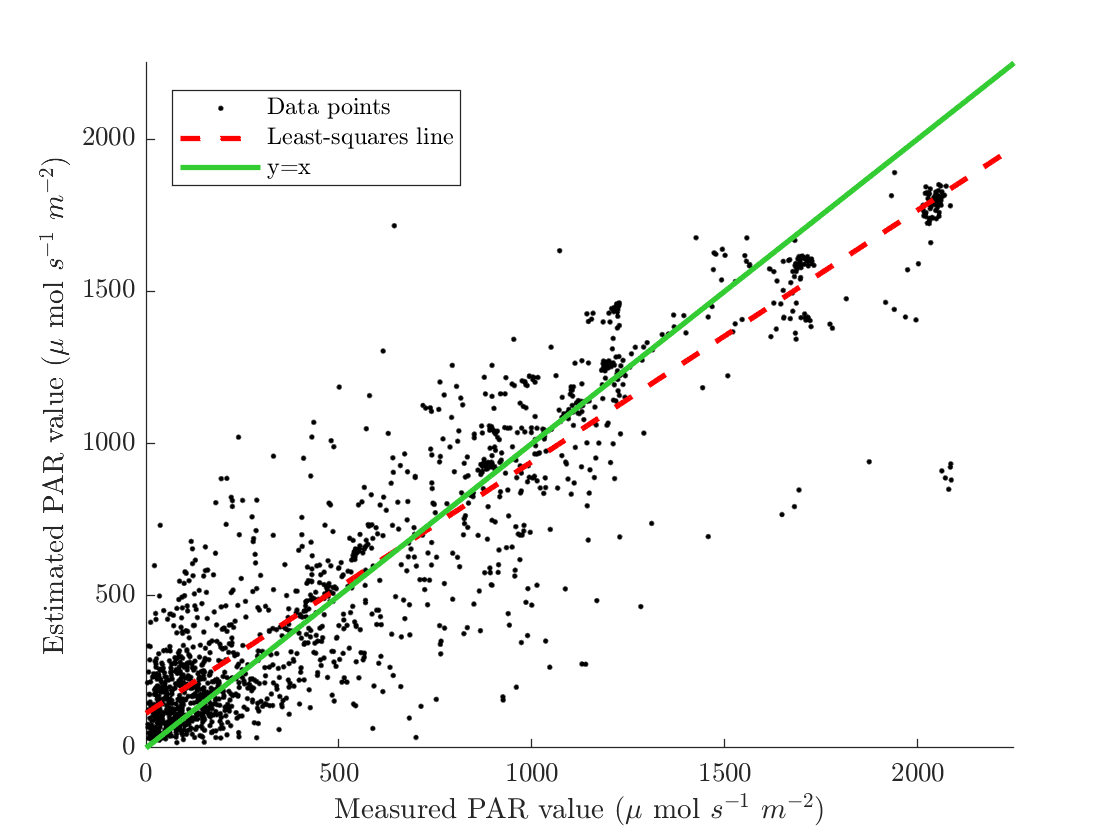}
    \label{fig:results-quantitative-1x1}}\hfill
    \subfloat[4 ceptometers per data point]{%
    \includegraphics[width=\linewidth]{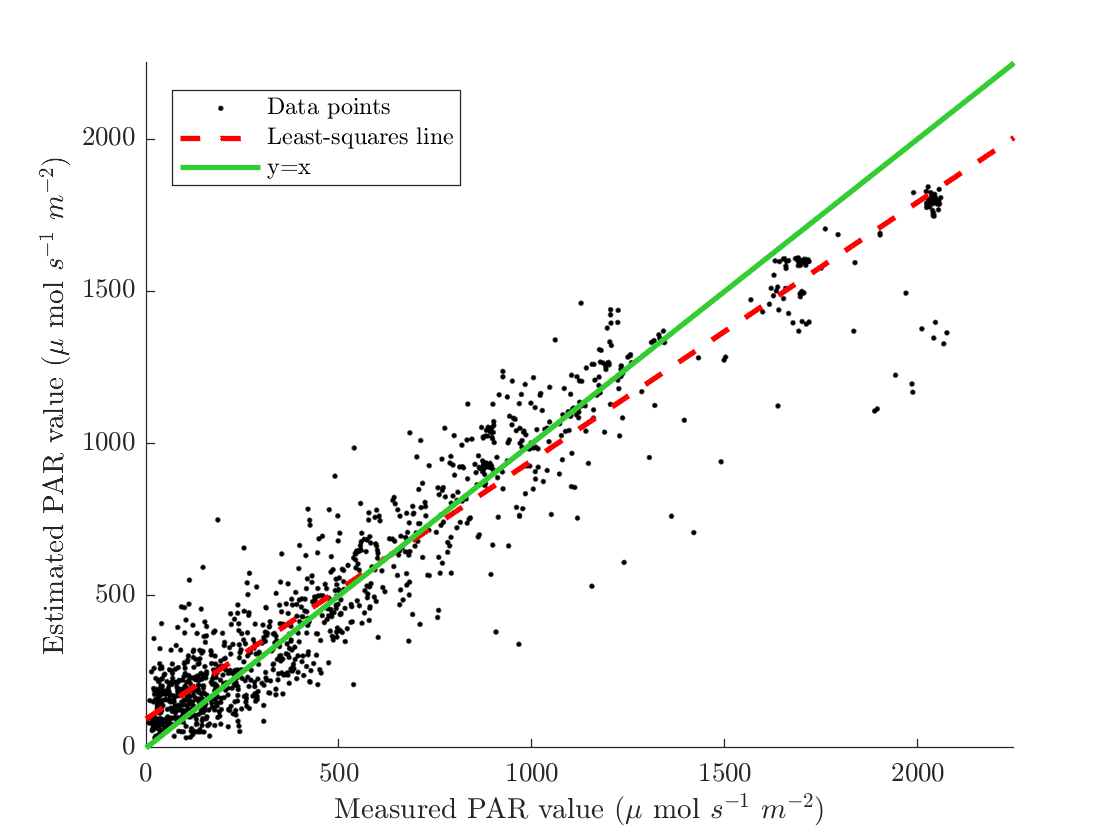}
    \label{fig:results-quantitative-2x2}}
    \caption{Estimated vs measured ceptometer readings.  Experiment performed with model parameters $s_{vox} = 0.1m$, $w_{vox} = 1$, $S=19$ and $\beta = 0.80$.  The green line plots y=x while the red dashed line represents the line of best fit.  (a) presents with $R^2 = 0.854$, slope $m = 0.826$ and $RMSE = 237 \mu mol s^{-1} m^{-2}$, while (b) presents with $R^2 = 0.923$, slope $m = 0.849$ and $RMSE = 157 \mu mol s^{-1} m^{-2}$}
    \label{fig:results-quantitative}
\end{figure}

The scatter plot in Figure~\ref{fig:results-quantitative-1x1} shows the results of comparing estimated energy values with ceptometer measurements across all available data sets when estimated using the optimal parameters determined here.  The green line plots y=x which is the desired relationship between estimated and measured values.  Meanwhile the red dashed line represents the line of best fit, demonstrating a strong relationship with $R^2 = 0.854$, $m = 0.826$ and $RMSE = 237 \mu mol s^{-1} m^{-2}$.

Several clusters are apparent within the scatter plot. The large cluster near the origin was where the model and ceptometer agreed on heavy shade. Several smaller clusters occur along the line of equality for higher energy values in full sun, which appear in smaller clusters than the dark points since every dataset represents a different time of day on different days, so the maximum light available varied.  The noise in the model is most evident in the intermediate regions of the graph where the ground under the canopy is in partial shade. In these cases, the high spatial frequency of dappled light and shade patterns challenged the spatial resolution of the model.  The plot in Figure~\ref{fig:results-quantitative-2x2} illustrates the result when this dappling was compensated for, by averaging over ceptometer readings and model estimates in a square sliding window of size 2m.  This resulted in a strengthened relationship ($m=0.849$, $R^2 = 0.923$) and reduced noise ($RMSE=157$).

\subsection{Parameter Optimisation}
\label{sec:results-parameters}

\begin{figure}
    \centering
  \subfloat[Gradient $m$]{%
       \includegraphics[width=0.95\linewidth]{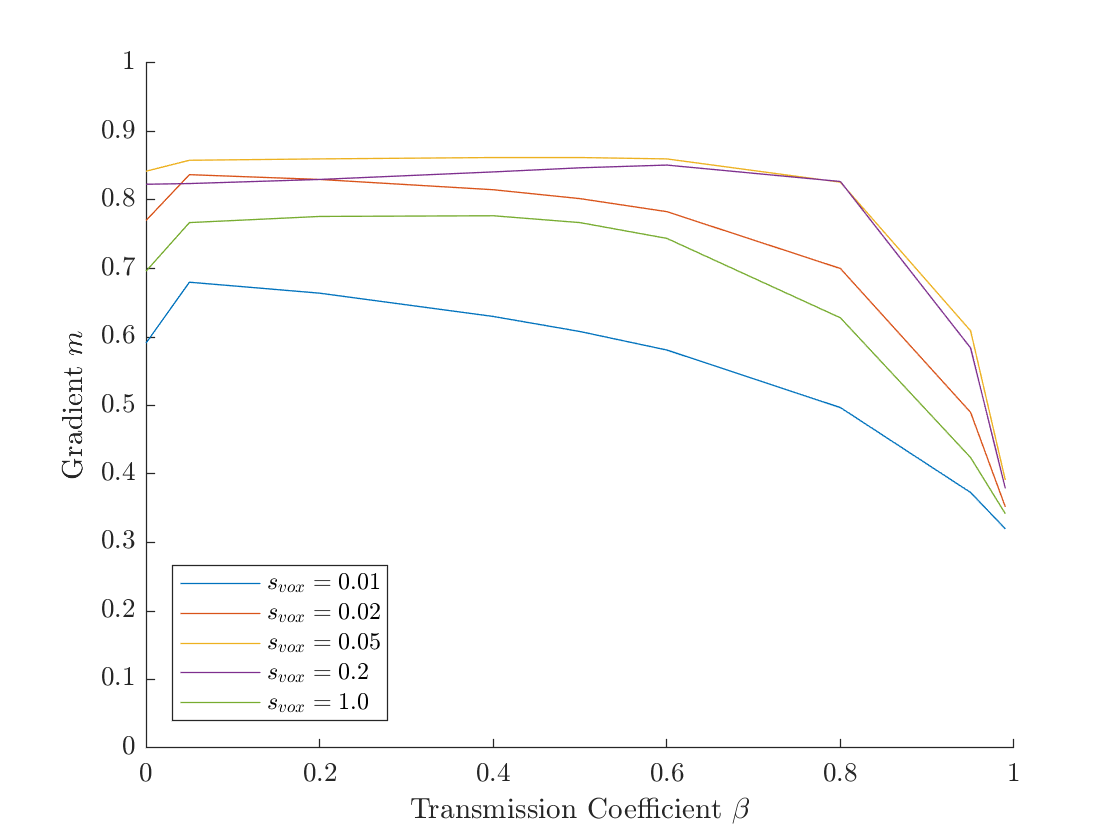}
    \label{fig:tuning-beta-m}}\hfill
  \subfloat[$R^2$]{%
       \includegraphics[width=0.95\linewidth]{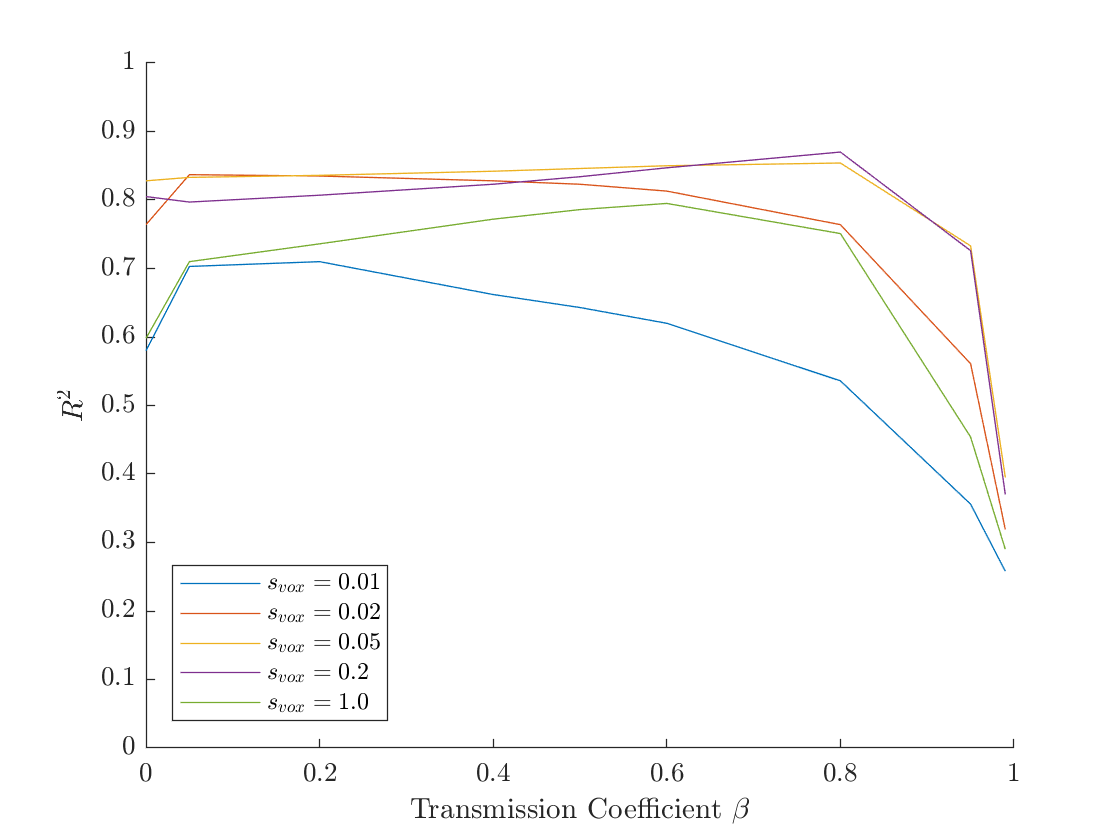}
    \label{fig:tuning-beta-r2}}\hfill
  \subfloat[$RMSE$]{%
       \includegraphics[width=0.95\linewidth]{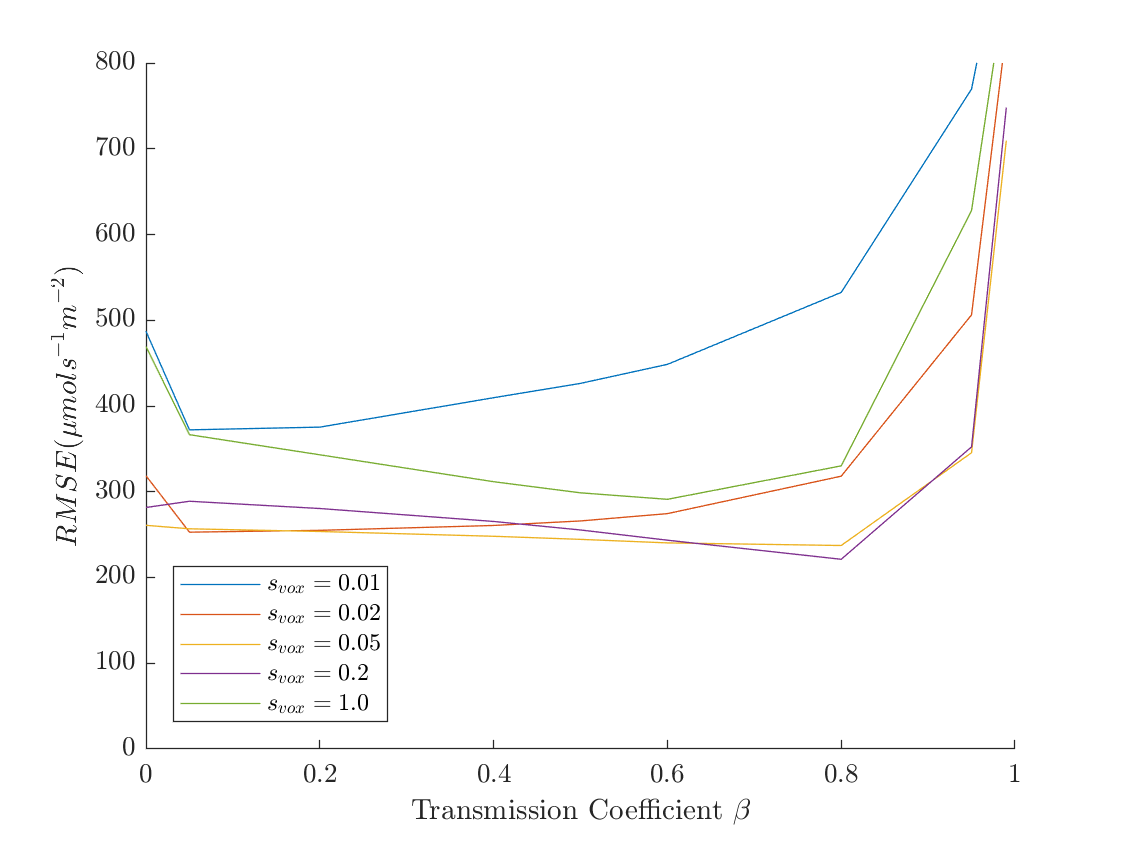}%
    \label{fig:tuning-beta-rmse}}\hfill
    \caption{Results when changing $\beta_{f}$.  $w_{vox}=1$ and $S=19$ are fixed while $s_{vox}$ is varied as shown.}
    \label{fig:results-tuning-beta}
\end{figure}

The parameters in Table~\ref{tab:method-tuning} form a multi-dimensional optimisation problem with a significant runtime per step prohibiting a joint optimisation, therefore a grid search was performed starting with the two most influential parameters, foliage transmission coefficient $\beta$ and voxel resolution $s_{vox}$, before fixing those and varying the others.  Figure~\ref{fig:results-tuning-beta} shows performance of the model plotted for varying $\beta$, with a separate line for different voxel sizes $s_{vox}$.  The results deteriorate with voxel sizes too large (0.5m) and too small (0.01m).  Across most voxel sizes a transmission coefficient of $\beta = 0.8$ was optimal.

\begin{figure}
	\centering
  \subfloat[Plot]{%
        \includegraphics[width=\linewidth]{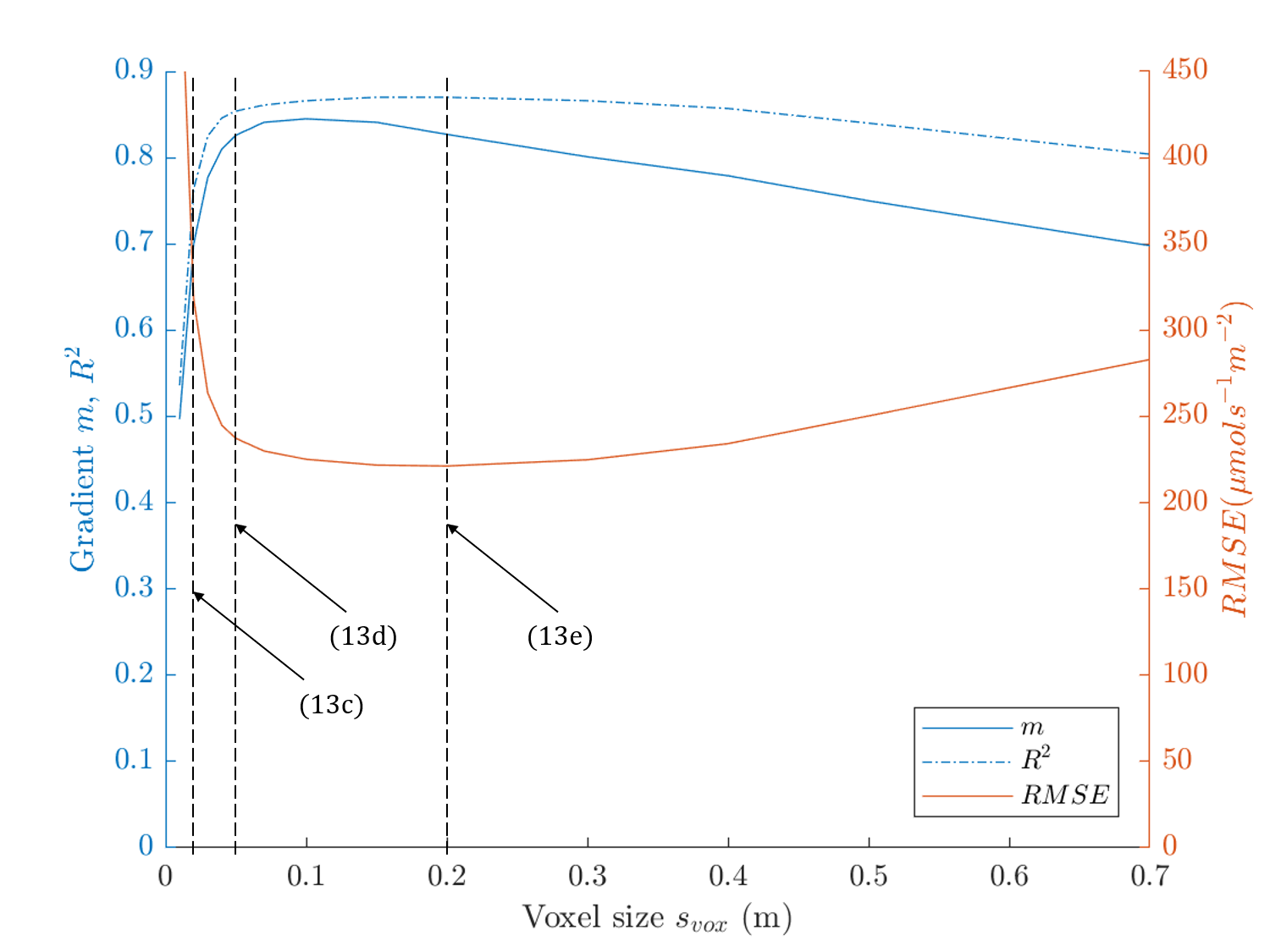}}\hfill
  \subfloat[Photo]{%
        \includegraphics[width=0.45\linewidth]{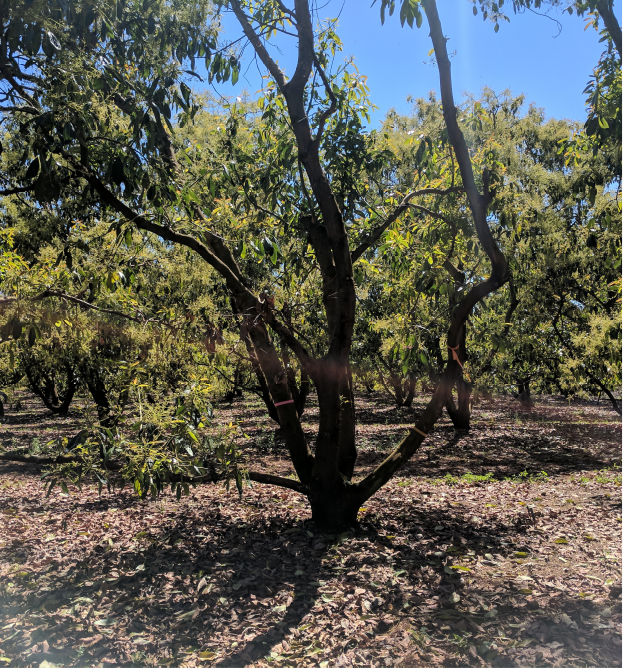}
        \label{fig:results-qualitative-photo}}\hfill
  \subfloat[$v_{size}=0.02m$]{%
        \includegraphics[width=0.45\linewidth]{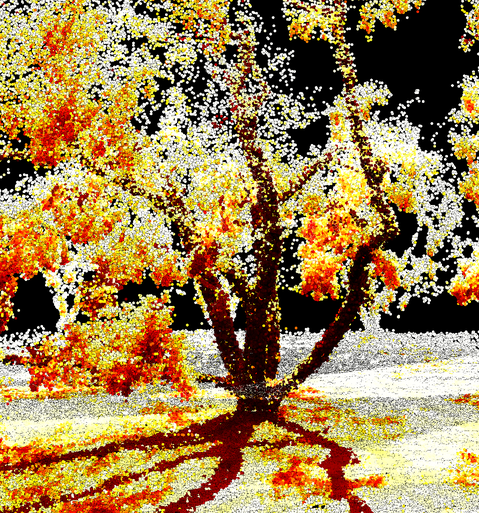}
        \label{fig:results-qualitative-1}}\hfill
  \subfloat[$v_{size}=0.05m$]{%
        \includegraphics[width=0.45\linewidth]{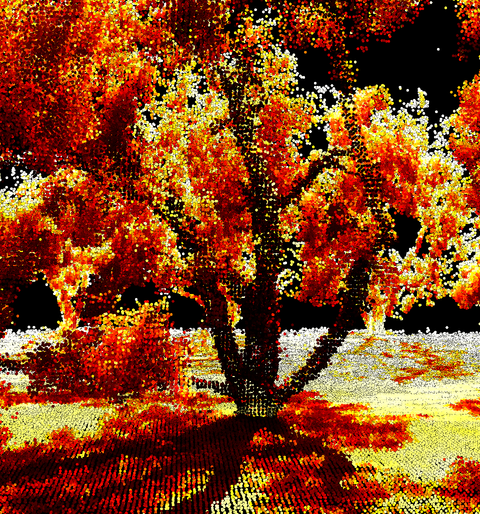}
        \label{fig:results-qualitative-2}}\hfill
  \subfloat[$v_{size}=0.20m$]{%
        \includegraphics[width=0.45\linewidth]{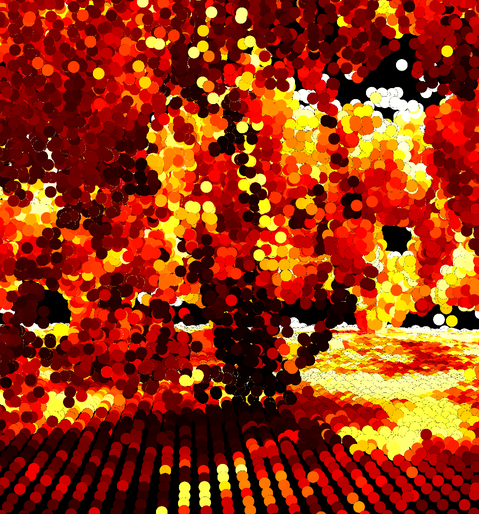}
        \label{fig:results-qualitative-3}}\hfill
	\caption{Effect on results when $\beta_f$ is fixed at 0.80, and $s_{vox}$ is varied.  Other parameters are fixed at $S=19$ and $w_{vox}=1$.}
	\label{fig:results-tuning-vs}
\end{figure}

With $\beta$ fixed, Figure~\ref{fig:results-tuning-vs} plots the model performance for different voxel sizes at a higher resolution than Figure~\ref{fig:results-tuning-beta}.  The performance degraded at the upper and lower ranges, but there was no clear peak or optimal value.  The slope demonstrated the most informative peak, suggesting an acceptable parameter range of $s_{vox} \in [0.04,0.1] m$.  The data suggested any voxel size in this range at $\beta=0.8$ would provide near-optimal results.  

Qualitative comparisons of shadow quality between models at different voxel sizes and the photo in Figure~\ref{fig:results-qualitative-photo} were performed and displayed in Figure~\ref{fig:results-tuning-vs}.   Small voxel sizes ($<0.03m$) demonstrated an overly distinct shadow as seen in Figure~\ref{fig:results-qualitative-1}, while larger voxel sizes ($>0.07m$) deteriorated recognisable features as seen in Figure~\ref{fig:results-qualitative-3}, despite comparable $R^2$ and $RMSE$. It was determined that $s_{vox} = 0.05m$ provided a good balance of qualitative and quantitative performance.  

While the plot in Figure~\ref{fig:results-tuning-vs} was generated with a fixed $\beta = 0.80$, we also explored the results at other values of $\beta$ to compare voxel sizes at their locally optimal transmission coefficient rather than a single global optimum.  These results were excluded for brevity, but demonstrate the same trends.

\begin{figure}
	\centering
	\includegraphics[width=0.95\linewidth]{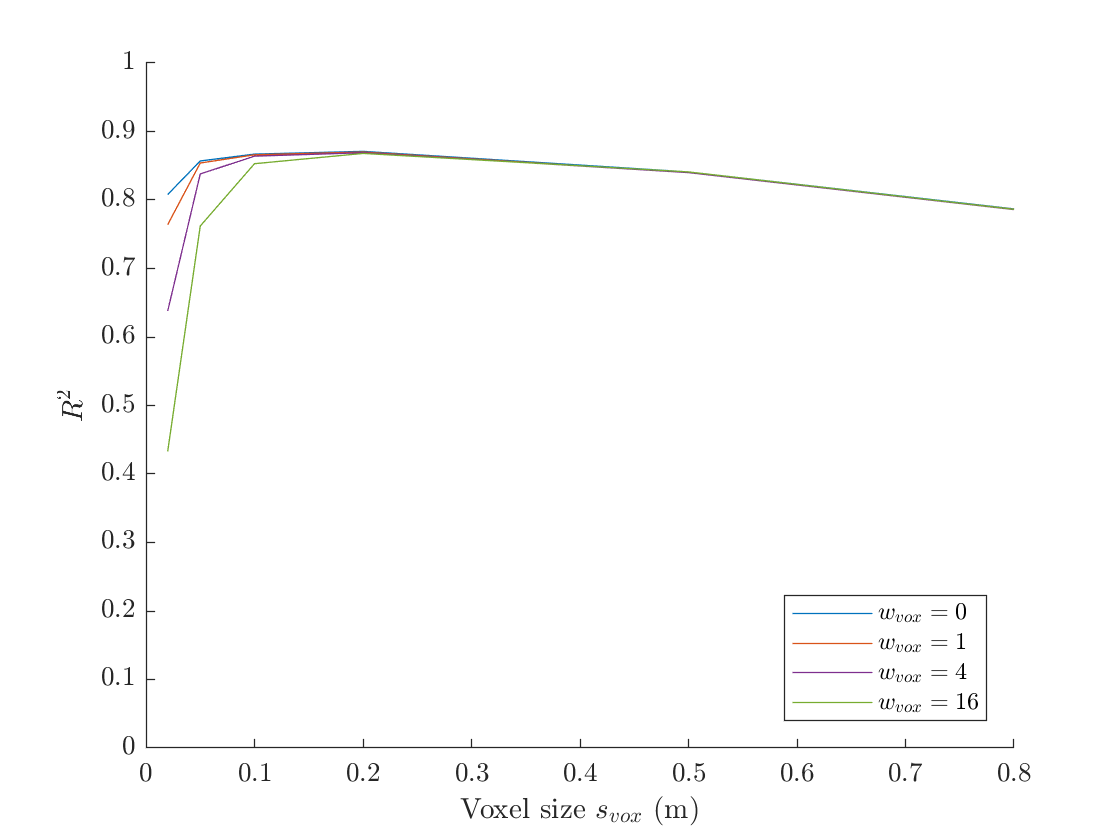}
	\caption{Effect on results when $w_{vox}$ is varied.  Other parameters are fixed at $S=19$, $\beta=0.80$ and $s_{vox}=0.1$}
	\label{fig:results-tuning-vw}
\end{figure}

Once optimal values were selected for $\beta=0.8$ and $s_{vox}=0.05$, we investigated the effect of changing $w_{vox}$ (the minimum number of LiDAR points in a voxel for it to qualify as solid matter).  This parameter is related to $s_{vox}$ and used primarily for noise rejection.  Figure~\ref{fig:results-tuning-vw} shows the $R^2$ against $s_{vox}$ for different values of $w_{vox}$ to illustrate the relationship between the two parameters.  $w_{vox} = 0$ never performed worse than any other value, suggesting that noise filtering was not required for our data.  Further, When the weight was made too large compared to the voxel size, performance was reduced as critical details were removed from the point cloud with excessive noise reduction.  Once the size-to-weight ratio was sufficiently large, the weight used appeared to have no quantitative effect.

\begin{table}[h]
    \centering
    \begin{tabular}{cccccccc}
        		 & \multicolumn{3}{c}{With sun node} & \multicolumn{3}{c}{Without sun node} &\\
         $S$ & $m$ & $RMSE$ & $R^2$ & $m$ & $RMSE$ & $R^2$ & Runtime \\
         \hline
         19 & 0.845 & 225 & 0.866 & 0.732 & 461 & 0.606 & 1m53s \\
         121 & 0.847 & 224 & 0.867 & 0.826 & 275 & 0.812 & 15h28m6s\\
         315 & 0.847 & 224 & 0.867 & 0.841 & 252 & 0.837 & 39h38m8s \\
         841 & 0.847 & 224 & 0.867 & 0.843 & 239 & 0.851 & 106h13m8s \\
         1271 & 0.847 & 224 & 0.867 & 0.843 & 247 & 0.843 & 159h46m48s \\
         1983 & 0.847 & 224 & 0.867 & 0.847 & 233 & 0.858 & 249h24m30s
    \end{tabular}
    \caption{Model error metrics for different sky complexities.  Experiments were performed with constant model parameters $s_{vox} = 0.1m$, $w_{vox} = 1$ and $\beta = 0.80$. Runtime was measured as the CPU time used by the whole process.  The model was parallelised across 8 cores, so the real runtime was approximately 1/8th the reported. }
    \label{tab:results-tuning-sky}
\end{table}

Finally the sky resolution $S$ was varied, both with and without the use of a dedicated sky-node for the exact sun position, with resulting slope, $R^2$ and $RMSE$ reported in Table~\ref{tab:results-tuning-sky}.  This shows that the resolution of the sky is generally unimportant if a dedicated sun node is used.  Without a dedicated node, lower resolutions suffered performance losses.  A significant difference in runtime was also demonstrated.  For this parameter, we used an $s_{vox}$ value of 0.1m to reduce model runtime, but as shown in Figure~\ref{fig:results-tuning-vs} this performed equally well as the chosen voxel size of 0.05m.

\begin{figure}
    \centering
        \includegraphics[width=0.95\linewidth]{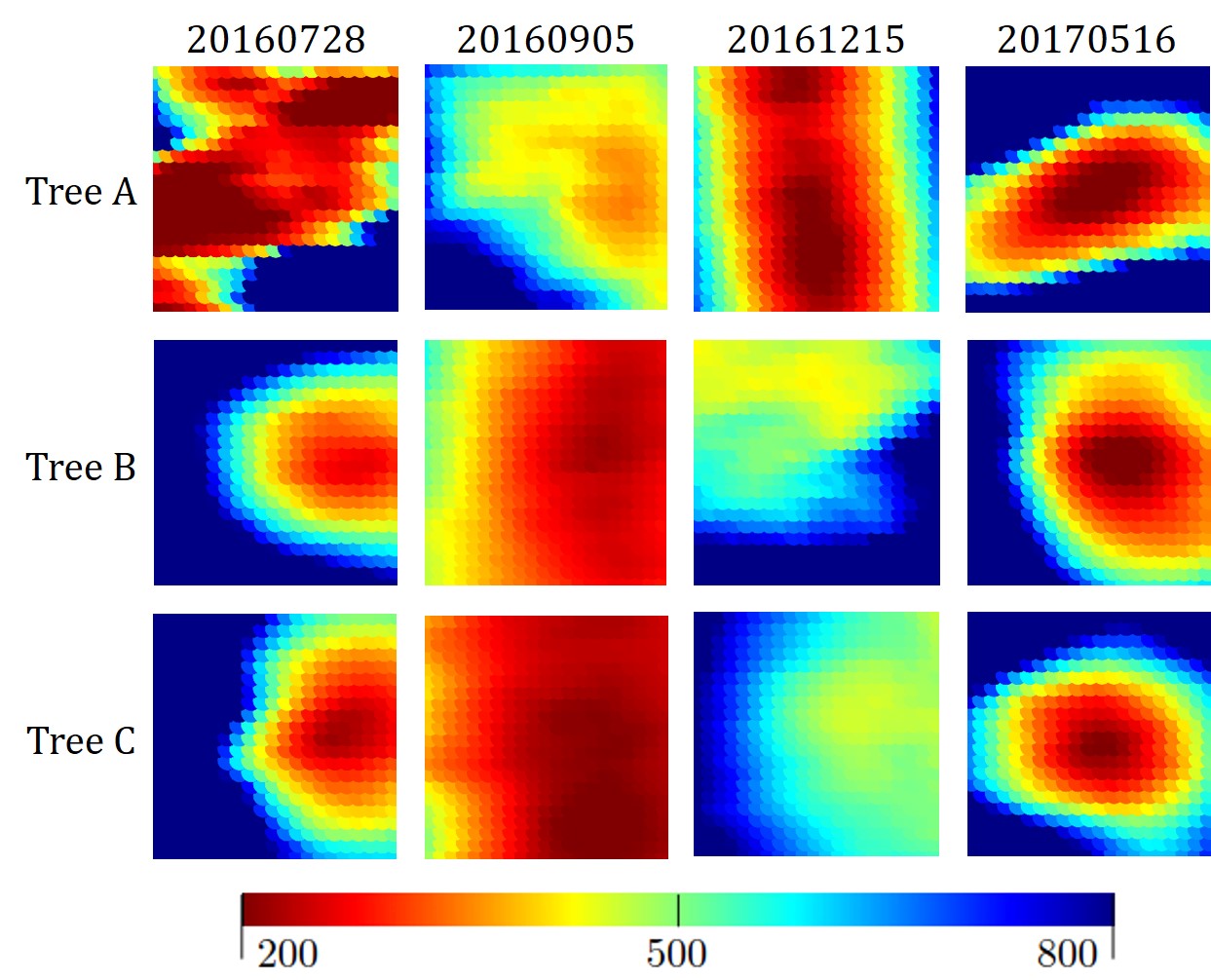}
  \caption{Heat maps of $RMSE$ across offsets for $\Delta x,\Delta y \in [-1,1]$ relative to $[0,0]$ at the centre which represents the original manually generated ceptometer placement.  Each individual LiDAR scan was represented here and all offset profiles use the same colour scale.}
  \label{fig:offset-optimisation} 
\end{figure}

Figure~\ref{fig:offset-optimisation} shows the performance of the model as a heat-map of RMSE for different $[\Delta x,\Delta y] \in [-1,1]m$ offsets for all data sets.  The offset is relative to the manually selected ceptometer position at $[\Delta x,\Delta y]= [0,0]$.  The response to these offsets demonstrated the sensitivity of precise shadow location on the measurement grid.  Some trees showed a clear optimal offset while others had a noticeable bias in X but not in Y.

\begin{table}[h]
    \centering
    \begin{tabular}{cccc}
        Experiment & $m$ & $RMSE$ & $R^2$ \\
        \hline
        Basic & 0.845 & 225 & 0.866 \\
        Incorrect point cloud & 0.246 & 2050 & 0.0726 \\
        Incorrect time & 0.663 & 504 & 0.564 \\
        Incorrect date & 0.558 & 607 & 0.471 \\
        No dedicated sun node & 0.732 & 461 & 0.606 \\
        No diffuse light & 0.911 & 251 & 0.839 
    \end{tabular}
    \caption{Degradation of error metrics when intentionally mismatching data for validation experiments.  All of these experiments were performed using model parameters $w_{vox} = 0.1m$, $w_{vox} = 1$, $S = 19$ and $\beta_{f} = 0.80$.}
    \label{tab:results-validation-incorrects}
\end{table}

 Table~\ref{tab:results-validation-incorrects} demonstrates the difference in results when different aspects of the model were purposefully altered.  Without any alterations, an $R^2$ of 0.866 was achieved, with an RMSE of 225.  The incorrect time and date worsened the results with $R^2 = 0.564$ and $R^2 = 0.471$ respectively, while removing the sun node caused a smaller deterioration in the relationship ($R^2 = 0.606$) but the error more than doubled ($RMSE=461$).  As was demonstrated in Table~\ref{tab:results-tuning-sky}, this would likely be far more significant with a smaller sky resolution but is almost unnoticeable at higher resolutions.  Removing the diffuse light component caused only a slight reduction in performance ($R^2 = 0.839$), though we would expect it to be worse in data sets during cloudy days with a higher diffuse fraction.  Use of the point cloud from an incorrect tree destroyed the relationship, with an $R^2$ of 0.0726, implying that modelling the geometric characteristics of the tree and its neighbours (for instance, height, volume, density and specific geometries) was critical.


\begin{figure}
    \centering
    \includegraphics[width=\linewidth]{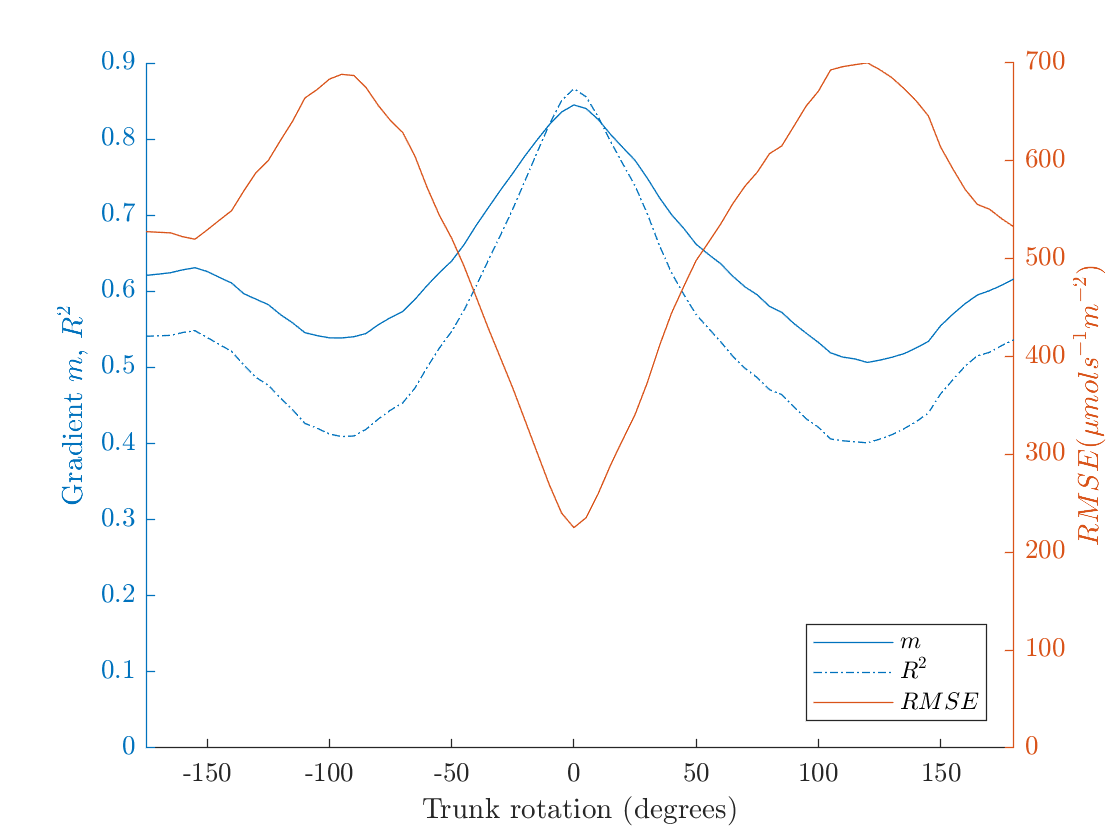}
    \caption{Error measures of model as tree trunks were rotated from 0 to 360 degrees.  Comparison performed using model parameters $s_{vox} = 0.1m$, $w_{vox} = 1$, $S = 19$ and $\beta_{f} = 0.80$.}
    \label{fig:results-validate-rot}
\end{figure}

The results of artificially rotating trees from their true alignment were shown in Figure~\ref{fig:results-validate-rot}, which revealed that the model performed best for the correct alignment and was sensitive to changes in alignment with a clear drop in performance for angular errors as small as 5 degrees.  Upon rotation, the tree maintained its large-scale features such as volume and height, while specific intra-tree geometry was varied.  This showed that it would be insufficient to model the tree in a simplistic fashion, for instance using a generic tree shape with the same characteristics (such as height) but none of the specific geometric detail.

\section{Discussion} 

The results presented in Section~\ref{sec:results} validated the model's energy estimates against measurements taken on the canopy floor, whereas the intended use of the model in future work is to estimate the energy distribution throughout the entire 3D canopies (as seen in Figure~\ref{fig:results-tuning-vs}). It was infeasible to gather ceptometer measurements distributed in 3D, and so the 2D canopy floor validation serves as a reasonable proxy. We believe this is valid because although the spatial arrangement of ceptometer data is two dimensional, the ray paths intersect the full complexity of 3D geometry prior to the planar intersection, and the importance of the specific 3D geometry was demonstrated. 

Validation was performed for instantaneous measurements because the cost of ceptometer sensors prohibited leaving one per node in the field. The model, however, permits estimation of energy absorption and distribution across time spans such as an entire growing season by integrating the changing sky throughout the duration of the simulation.  Figure~\ref{fig:composite-sky} demonstrated that as the sky is integrated, the variations in the diffuse component of light were averaged out.  This means short-term effects like local cloud movements which are unpredictable but may have introduced errors in the ceptometer validation in this study were unlikely to have a significant impact on the composite sky.  The more critical value to model in this case was the total diffuse light, which was extrapolated from public record and does not require sensors on location.  

Introducing a dedicated sky node demonstrated a practical improvement in runtime for instantaneous estimates, since the direct light from the sun was dominant at any one given time and the diffuse light distributed across the remainder of the sky was unlikely to provide a significant effect on individual ceptometer readings, as shown in Table~\ref{tab:results-validation-incorrects}.  However, if this approach were used for time spans, the resolution of the sky would be inflated by the addition of many continuously placed nodes, which would prove infeasible beyond a certain length of time.  For this reason, a sky resolution should be chosen which can match the instantaneous performance of the dedicated sun node without overly compromising runtime.  The results reported in Table~\ref{tab:results-tuning-sky} suggested a resolution of 841 sky nodes would be appropriate.

Furthermore, the geometry of the neighbouring trees gains an increased significance when the energy estimate includes the earlier and later hours of the day, as the direct light from the sun is more likely to pass through neighbours on its path to the tree of interest.  In our data, trees in neighbouring rows were insufficiently scanned by LiDAR, such that the shadows cast by these trees were not able to be modelled.  While this did not affect the ceptometer measurements used in validation, it may impact seasonal light estimates so collecting comprehensive LiDAR scans of neighbouring trees would improve performance.  These scans could be provided by more extensive use of the Zebedee handheld LiDAR, or by other mobile LiDAR systems like that presented by~\cite{underwood2016mapping} and \cite{stein2016image}, which are designed to capture per-tree geometry at the scale of the whole orchard.  

If a mobile platform were used to scan entire orchard blocks, there would be no concern of unmodelled neighbours.  As mentioned, the model would require a high-resolution sky to achieve an appropriate level of accuracy, and with a larger point cloud, the run time for the model would increase significantly.  However, the model is highly parallelisable as the ray traced from each sky node can be processed in isolation.

\subsection{Future work}

The current model takes into account a classification of LiDAR points as trunk or foliage, which was provided by manual labelling for this study.  Existing algorithms for branch/foliage classification will be explored, to provide a completely automated end-to-end solution in future work.

Further, since the ultimate aim of the model is to provide data that can help improve orchard yield, the predictive power of the model's output for estimating yield and fruit size will be tested and validated.  While yield is dependent on a variety of factors that are not captured in the model, a proven correlation between estimated light intake and yield will enable the model to be applied to pruning recommendation towards automated orchard decision support and management.  

Using the model described here and a digitised orchard (e.g. using a mobile LiDAR platform), existing pruning methods could be evaluated with the aim of maximising the total light or light distribution of fruit trees.  Further, such a system could suggest variations on traditional methods, which are more optimised for this purpose.

\section{Conclusion}

A model for simulating the light energy captured by individual fruit trees in an orchard was presented.  The sky was estimated at several particular times and places, taking weather into account, and the light from each part of the sky was traced through a tree model created using a hand-held LiDAR. 

The model was validated and optimised using ceptometer measurements captured in parallel with the LiDAR scans.  Strong agreement was observed between the model and ceptometer data on the canopy floor ($R^2 = 0.854$, $RMSE = 237 \mu mol s^{-1} m^{-2}$).  An additional validation was performed to assess the importance of each major sub-component, which demonstrated the overall complexity of the algorithm was justified. 

The validated model is suitable for further development towards an orchard decision support system for pruning.

\subsubsection*{Acknowledgements}
This work is supported by the Australian Centre for Field Robotics (ACFR) at The University of Sydney and by funding from the Australian Government Department of Agriculture and Water Resources as part of its Rural R\&D for profit programme.  Thanks to Sushil Pandey, Nicholas Anderson and Kerry Walsh from Central Queensland University for providing the ceptometer data, as well as Chad Simpson, Chris Searle and Simpson Farms for their support and to Andrew Robson from the University of New England for selecting target trees. Thanks also to Neil White for the ongoing discussions and collaboration. Finally, thanks to Vsevolod (Seva) Vlaskine and the software team for their continuous support, and to Salah Sukkerieh for his continuous direction and support of Agriculture Robotics at the ACFR. For more information about robots and systems for agriculture at the ACFR, please visit http://sydney.edu.au/acfr/agriculture.

\subsubsection*{Conflicts of interest}
The authors declare no conflict of interest.  The founding sponsors had no role in the design of the study; in the collection, analyses or interpretation of the data; in the writing of the manuscript; nor in the decision to publish the results.

\bibliography{main}{}
\end{document}